\documentclass[fleqn,usenatbib]{mnras}

\usepackage{newtxtext,newtxmath}

\usepackage[T1]{fontenc}
\usepackage{ae,aecompl}
\usepackage{lscape}

\usepackage{graphicx}	
\usepackage{amsmath}	

\newcommand{\NGTS}{NGTS}

\newcommand{\kms}{km\,s$^{-1}$}
\newcommand{\ms}{m\,s$^{-1}$}
\newcommand{\masy}{mas\,y$^{-1}$}
\newcommand{\mpl}{\mbox{M$_{p}$}}
\newcommand{\rpl}{\mbox{R$_{p}$}}
\newcommand{\mstar}{\mbox{M$_{s}$}}
\newcommand{\rstar}{\mbox{R$_{s}$}}
\newcommand{\mjup}{\mbox{M$_{\rm J}$}}
\newcommand{\rjup}{\mbox{R$_{\rm J}$}}
\newcommand{\msun}{\mbox{M$_{\odot}$}}
\newcommand{\rsun}{\mbox{R$_{\odot}$}}

\newcommand{\gccc}{g\,cm$^{-3}$}
\newcommand{\ergscm}{erg\,s$^{-1}$cm$^{-2}$}
\newcommand{\teff}{$T_{\rm eff}$}
\newcommand{\logg}{$\log g$}
\newcommand{\met}{\mbox{[Fe/H]}}
\newcommand{\age}{\mbox{t$_{s}$}}
\newcommand{\Nstar}{NGTS-33}

\newcommand{\Nplanet}{NGTS-33b}

\title[\Nplanet]{\Nplanet: A Young Super-Jupiter Hosted by a Fast Rotating Massive Hot Star}

\author[Alves, D. et al.]{
\parbox{\textwidth}{
Douglas R. Alves,$^{1,3}$\thanks{E-mail: \href{douglasalvesastro12@gmail.com}{douglasalvesastro12@gmail.com}\\
\href{dalves@das.uchile.cl}{dalves@das.uchile.cl}
}
James S. Jenkins,$^{2,3}$
Jose I. Vines,$^{4}$
Matthew~P.~Battley,$^{9}$
Monika Lendl,$^{9}$
Fran\c{c}ois Bouchy,$^{9}$
Louise D. Nielsen,$^{10}$
Samuel Gill,$^{5,6}$
Maximiliano~Moyano,$^{4}$
D.~R.~Anderson,$^{4}$
Matthew R. Burleigh,$^{11}$
 Sarah L. Casewell,$^{11}$
Michael R.~Goad,$^{11}$
Faith~Hawthorn,$^{5,6}$
Alicia Kendall,$^{11}$
James~McCormac,$^{5,6}$
Ares~Osborn,$^{8}$
Alexis~M.~S.~Smith,$^{7}$
St\'{e}phane~Udry,$^{9}$
Peter~J.~Wheatley,$^{5,6}$
Suman Saha,$^{2,3}$
L\'{e}na~Parc,$^{9}$
Arianna~Nigioni,$^{9}$
Ioannis~Apergis,$^{5,6}$
Gavin Ramsay$^{12}$}
\vspace{3mm}
\\
$^1$Departamento de Astronom\'ia, Universidad de Chile, Casilla 36-D, Santiago, Chile\\
$^2$Instituto de Estudios Astrof\'isicos, Universidad Diego Portales, Av. Ej\'ercito 441, Santiago, Chile\\
$^3$Centro de Astrof\'isica y Tecnolog\'ias Afines (CATA), Casilla 36-D, Santiago, Chile\\
$^4$Instituto de Astronom\'ia, Universidad Cat\'olica del Norte, Angamos 0610, 1270709, Antofagasta, Chile\\
$^{5}$ Department of Physics, University of Warwick, Gibbet Hill Road, Coventry CV4 7AL, UK \\
$^{6}$ Centre for Exoplanets and Habitability, University of Warwick, Gibbet Hill Road, Coventry CV4 7AL, UK\\
$^{7}$Department of Extrasolar Planets and Atmospheres, Institute of Planetary Research, German Aerospace Center (DLR), Rutherfordstra\ss e 2, 12489 Berlin, Germany\\
$^{8}$Department of Physics and Astronomy, McMaster University, 1280 Main St W, Hamilton, ON, L8S 4L8, Canada\\ 
$^{9}$ Departement d'Astronomie, Universit\'e de Gen\`eve, 51 chemin Pegasi, 1290 Sauverny, Switzerland\\
$^{10}$ University Observatory, Faculty of Physics, Ludwig-Maximilians-Universit{\"a}t M{\"u}nchen, Scheinerstr. 1, 81679 Munich, Germany\\
$^{11}$ School of Physics and Astronomy, University of Leicester, Leicester LE1 7RH, UK\\
$^{12}$Armagh Observatory and Planetarium, College Hill, Armagh, BT61 9DG, UK}
\date{}

\pubyear{2024}

\begin{document}
\label{firstpage}
\pagerange{\pageref{firstpage}--\pageref{lastpage}}
\maketitle

\begin{abstract}
In the last few decades planet search surveys have been focusing on solar type stars, and only recently the high-mass regimes. This is mostly due to challenges arising from the lack of instrumental precision, and more importantly, the inherent active nature of fast rotating massive stars. Here we report \Nplanet~(TOI-6442b), a super-Jupiter planet with mass, radius and orbital period of 3.6 $\pm$ 0.3 \mjup, 1.64 $\pm$ 0.07 \rjup~and $2.827972 \pm 0.000001$ days, respectively. The host is a fast rotating ($0.6654 \pm 0.0006$ day) and hot (T$_{\rm eff}$ = 7437 $\pm$ 72 K) A9V type star, with a mass and radius of 1.60 $\pm$ 0.11 \msun~and 1.47 $\pm$ 0.06 \rsun, respectively. Planet structure and Gyrochronology models shows that \Nstar~is also very young with age limits of 10-50 Myr. In addition, membership analysis points towards the star being part of the Vela OB2 association, which has an age of $\sim$ 20-35 Myr, thus providing further evidences about the young nature of \Nstar. Its low bulk density of 0.19$\pm$0.03 \gccc~is 13$\%$ smaller than expected when compared to transiting hot Jupiters with similar masses. Such cannot be solely explained by its age, where an up to 15$\%$ inflated atmosphere is expected from planet structure models. Finally, we found that its emission spectroscopy metric is similar to JWST community targets, making the planet an interesting target for atmospheric follow-up. Therefore, \Nplanet's discovery will not only add to the scarce population of young, massive and hot Jupiters, but will also help place further strong constraints on current formation and evolution models for such planetary systems.

\end{abstract}
\clearpage
\section{Introduction}
\label{sec:intro}

The detection of the first hot Jupiters \citep[HJ;][]{mayor1995jupiter,charbonneau2000detection} set a great milestone in the exoplanet field, paving the way to the detection of several more extra-solar planets. Ground-based spectroscopic \citep{valenti2005spectroscopic, jenkins2009first} and photometric surveys \citep{bakos2004wide,pollacco2006wasp,2013EPJWC..4713002W} were initially responsible for the detection and characterisation of several massive HJs. Yet, though challenging, a few Neptunes \citep[e.g., HAT-P-11b, NGTS-14Ab;][]{bakos2010hat,smith2021ngts} have also been confirmed. Such selection bias towards massive planets is mostly due to the larger Doppler signal as well as transit depths induced by HJs given their mass and short orbital periods (P < 10 days), making their detection more favourable compared to less massive and/or longer period worlds. Besides that, ground-based missions are limited by atmospheric conditions \citep{cubillos2016correlated,o2022scintillation} as well as the day-night cycle, hence challenging the discovery of long-period planets. On the other hand, space-based missions such as CoRoT \citep{auvergne2009corot}, Kepler/K2 \citep{borucki2010kepler} and the Transiting Exoplanet Survey Satellite \citep[TESS;][]{TESS} were free of such atmospheric limitations, helping to reduce the bias by increasing the number of planet detections\footnote{5675 planets on 15 May 2024 according to \url{https://exoplanet.eu/}} significantly, with Kepler being responsible for the discovery of several (mini-)Neptunes (10M$_{\oplus}$$\leq$M$<$38M$_{\oplus}$) and (super-)Earths (1M$_{\oplus}$$\leq$M$<$10M$_{\oplus}$), whilst providing robust statistical constraints on planet occurrence rates \citep{mulders2015stellar, hsu2019occurrence}, as well as the multiplicity nature of rocky worlds \citep{cochran2011kepler,gillon2017seven}. In addition, the dearth of Neptune planets with orbital periods interior to 4 days, known as the Neptune desert, first dropped out of the Kepler data, and subsequently corroborated by the relatively low Neptune discoveries from TESS mission. Though rare, planets have been detected and characterised in this region \citep[e.g., LTT9779b;][]{jenkins2020ultrahot,hoyer2023extremely,fernandez2024survival}, which were fundamental to constrain the most likely scenario for these planets. In addition, the evolution of giant planets may even be associated to the emergence and form of the desert, where a combination of tidal migration with atmospheric photoevaporation \citep{lopez2013role, owen2017evaporation} may have been able to strip off the migrating giants' envelopes, giving rise to a population of Neptune-desert planets.

Like the Neptune desert planets, the HJ population remains subjected to intense scrutiny, especially the transiting HJs (THJs), where besides their masses, radii (R$_{\rm p}$) and bulk densities ($\rho_{\rm p}$) can be estimated at high precision. Moreover, atmospheric follow-up may help reveal key properties (e.g, day-night temperatures, albedos/reflectivity and atmosphere abundances), making them great test beds for probing giant planet formation and evolution histories. For instance, HJs are not expected to form in-situ but farther out in the disc where an initial $\sim10$ M$_\oplus$ protocore mass is fundamental to the onset of a runaway gas accretion process leading to the formation of giant planets \citep{pollack1996formation,alibert2005models,piso2014minimum}. In addition, HJs formation and disc migration must occur quickly ($\sim$ 5-10 Myr), with timescales being a function of the protostar mass as well as the disc properties, specially its mass and metallicity \citep[see, ][]{fortney2021hot}. In fact, it has been noted that stellar metallicity (\met) correlates with giant planet fraction (f$_{\rm p}$), indicating that such planets are formed more effectively around metal-rich stars \citep{gonzalez1997stellar,santos2001metal,fischer2005planet,2020MNRAS.491.4481O}. Moreover, an increase in \met~leads to a higher f$_{\rm p}$, providing further evidences that disc metallicity plays an important role in the formation of giant planets \citep{santerne2016sophie,buchhave2018jupiter,barbato2019gaps}. For instance, \citet{jenkins2017new} finds a correlation between giant hosts \met~and P, where the population of giants exterior to P$\leq$100 days are more metal-rich with a \met~mean difference $\sim0.16$ dex. Therefore, they point out that such correlation may indicate the initial location where giant planets formed in the protoplanetary disc. Finally, giant planets seem to be more abundant around more massive stars, where higher f$_{\rm p}$ are observed as a function of stellar mass \citep{johnson2010giant,reffert2015precise,jones2016four}. Such results point to a correlation between the central star and disc masses, where the former plays an important role in the type of planets that will be formed. In other words, an increase in the stellar mass leads to higher disc masses, hence more material with which to quickly form massive giant planets before the disc disperses. Therefore, these results are in agreement with the current findings of massive stars harbouring massive gas giant planets, particularly at short periods where the HJ population is located.

As mentioned, the evolution of giant planets may have been subjected to disc migration rather than in-situ formation. Further evidence against the latter include the detection of atmospheric escape in HJs \citep[e.g., HD209458b, HD 189733b, KELT-19b;][]{vidal2003extended,des2010evaporation,wyttenbach2020mass} and high obliquities of close-in giants, which may be explained by the planets' evolution through high eccentricity ($e$) migration by secular interactions with outer companions \citep{fortney2021hot,vick2023high}. In fact, a handful of hot hosts have massive HJs with non-zero obliquities (e.g., KELT-17b, MASCARA-1b, TOI-1431b), with some even retrograde \citep[HAT-P-14b;][]{winn2011orbital}, whereas cool stars tend to have HJs with low obliquities \citep{winn2011orbital}, thus showing that massive giant planets might have gone through multiple migration channels. 

Here, we report the detection of \Nplanet, the first NGTS discovery of a super-Jupiter hosted by a massive star. This detection will add to the small, but growing number of massive systems, that will help us to better understand the formation and evolution processes of massive planets around hot stars. The manuscript is organised as follows, in $\S$ \ref{sec:obs}, we present the photometry extraction from NGTS and TESS lightcurves, spectroscopic follow-up with FEROS, HARPS and CORALIE and their respective spectral line activity diagnosis. $\S$ \ref{sec:analysis} describes the data analysis, where we extract stellar parameters ($\S$ \ref{sub:stellar}), assess its age ($\S$ \ref{subsub:Agestimation}), perform a global modelling to derive the planetary parameters ($\S$ \ref{sub:globalmodeling}). Stellar rotation period, transit timing variation as well as dynamical stability were probed in $\S$ \ref{sub:rotation}, \ref{sub:TTVs}, and \ref{sub:dynamicalStability}, respectively. Finally, we discuss our results in $\S$ \ref{sec:discussion} and set out the conclusions in $\S$ \ref{sec:concl}.

\subsection{The Next Generation Transit Survey}
The Next Generation Transit Survey \citep*[NGTS;][]{Chazelas2012,McCormac2017,wheatley2018next} is located at the ESO Paranal Observatory in Chile, with the objective of detecting new transiting planetary systems. The consortium has 12 telescopes with 0.2 m diameter each, and individual fields of view of 8 deg$^2$, which combined provides a wide-field of 96 deg$^2$.  

The NGTS detectors host 2K$\times$2K pixels, with individual pixels measuring 13.5 $\mu$m, which corresponds to an on-sky size of 5 arcseconds/pixel, thus providing high sensitivity images over a wavelength domain between 520$-$890nm. This combination allows 150 ppm photometry to be obtained on bright stars (V$<$10~mags) for multi-camera observations, while for single telescope mode at 30 min cadence, a precision of 400 ppm is achievable \citep{bayliss2022high}. The project has been operational since February 2016, and over the past 8 years has so far acquired over 400 billion measurements of over 30 million stars. Within this treasure-trove of data, the NGTS has discovered 26 new planetary systems \citep*[e.g.,][]{bayliss2018ngts, bryant2020ngts,tilbrook2021ngts,bouchy2024ngts}, with more yet to be confirmed. A few of the highlights include the discovery of the Neptune desert planet NGTS-4b \citep{west2019ngts}, an ultra short period Jupiter NGTS-6b \citep{vines2019ngts}, the shortest period hot Jupiter NGTS-10b around a K5V star \citep{mccormac2020ngts}, an inflated HJ around a low-mass and metal-poor K dwarf \citep{alves2022ngts}, and a few warm planets \citep[e.g., NGTS-29b, NGTS-30b;][]{gill2024toi,battley2024ngts}.
\section{Observations}
\label{sec:obs}
Here we describe the photometry and spectroscopic data acquisition and reduction that lead to the discovery of \Nplanet\,. Table \ref{tab:obs}, \ref{tab:ngts}, and \ref{tab:rvs} show the NGTS and TESS observation dates and settings, a portion of the normalised photometry and radial velocity (RV) used in the analysis, respectively.
\subsection{\NGTS~ Photometry}
\label{sub:ngtsphot}
\begin{figure}
	\includegraphics[width=\columnwidth]{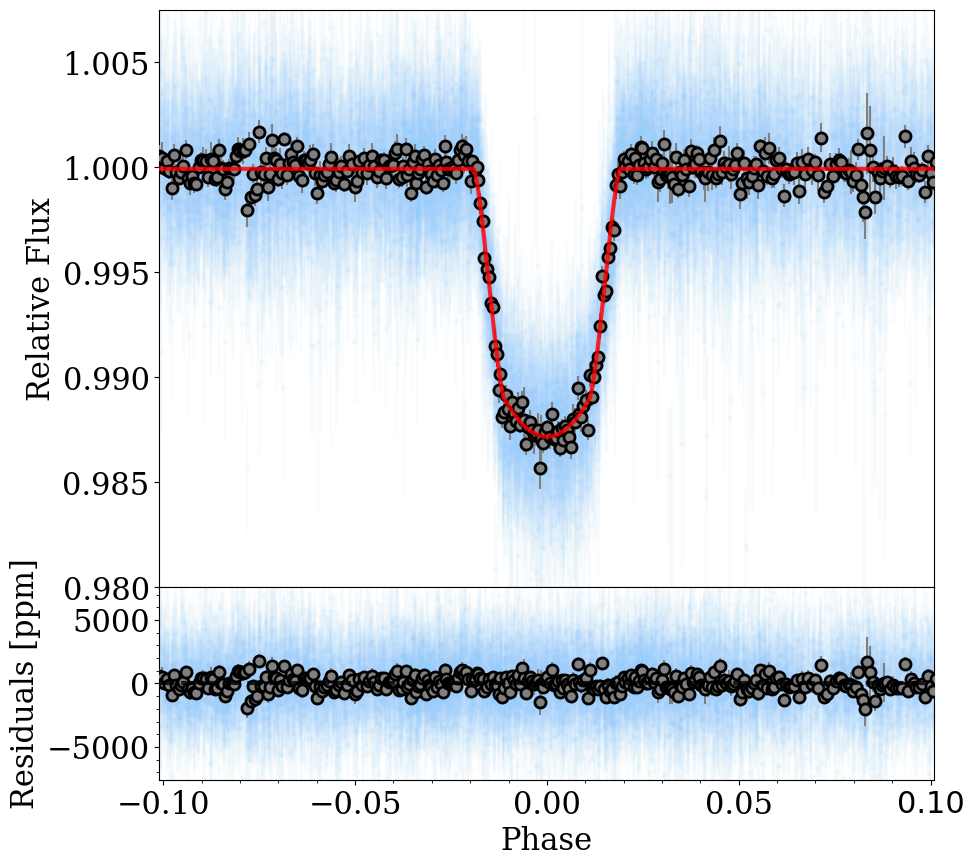}
    \caption{\textbf{Top}: NGTS detrended lightcurve phase-folded to the best-fitting period listed in Table \ref{tab:planet} and zoomed to show the transit event. Blue and black circles correspond to modelled photometric data and binned data with the associated photon noise error. The red line and shaded region show the median transit model and its 1-$\sigma$ confidence interval.
    \textbf{Bottom}: residuals to the best fit model.} 
    \label{fig:ngtsphot}
\end{figure}

\Nstar\ observations were taken during 2019-2021 in single camera mode and with 10 seconds exposure time per frame, as summarised in Table \ref{tab:obs}. Three independent telescopes were used at different epochs, yielding a total of 309,426 images from which stellar brightness was measured by aperture photometry with the CASUTools\footnote{\url{http://casu.ast.cam.ac.uk/surveys-projects/software-release}} package. As part of the data reduction, an adapted version of the SysRem algorithm \citep{Tamuz2005} and the box least-squares (BLS) fitting algorithm \citep[][]{kovacs2002box,collier2006fast} \texttt{ORION} were used to remove  nightly trends caused by atmospheric extinction and search for periodic transits in the timeseries, respectively. 28 transits in total were detected, from which seven had complete time coverage. A strong signal was detected at 2.83 days, and a validation process began in order to either confirm the signal as a transiting hot Jupiter or reject it as a false positive detection. For example, one of the vetting tests deal with background eclipsing binaries, where consecutive transits showing odd-even and/or V shaped morphologies could indicate candidates are false positives. \Nstar\ passed every validation step, and therefore further photometry and RV follow-up were obtained. Figure~\ref{fig:ngtsphot} shows the NGTS detection lightcurve wrapped around the best-fitting period $2.827969 \pm 0.000002$\,d computed from the global modelling ($\S$ \ref{sub:globalmodeling}). For a thorough description of the \NGTS~ mission, data reduction, and acquisition, we refer the reader to \citet{wheatley2018next}.
\begin{table}
\centering
\caption{NGTS and TESS observation settings}
\label{tab:obs}
\begin{tabular}{cccc}
&	NGTS Mission   &  & \\
Start date	&	End date       & nights & Camera\\
\hline
2019-Dec-18 & 2020-Mar-21 & 94 & 803 \\
2020-Sep-28 & 2021-Mar-25 & 178 & 804 \\
2021-Apr-26 & 2021-Jun-12 & 47 & 809 \\
\hline
&	TESS Mission   &  & \\
Start date	&	End date       & Sector & Camera\\
\hline
2019-Jan-07 & 2019-Feb-02 & 07 & 03 \\
2020-Dec-17 & 2021-Jan-13 & 33 & 03 \\
2021-Jan-13 & 2021-Feb-09 & 34 & 03 \\
2023-Jan-18 & 2023-Feb-12 & 61 & 03 \\
	\end{tabular}
\end{table}

\begin{figure*}
	\includegraphics[width=2\columnwidth,angle=0]{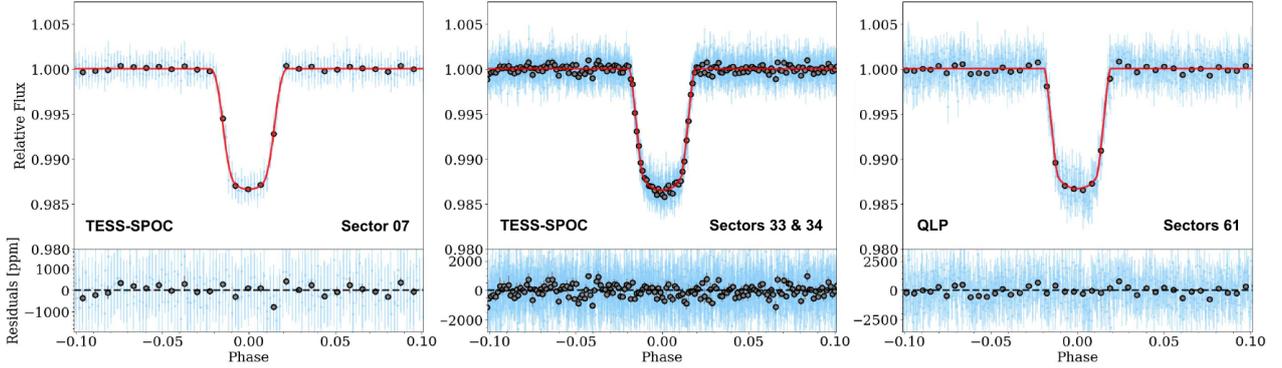}
    \caption{\textbf{Left}: Phase-folded, 30-minute detrended lightcurve from TESS-SPOC, Sector 07. \textbf{Center:} The same as the left plot but for Sectors 34 and 35 at 5-minute cadences. \textbf{Right:} Phase-folded, 3.33-minute detrended light curve from QLP, Sector 61. Colours and labels correspond to Fig. \ref{fig:ngtsphot}}
    \label{fig:tessphot}
\end{figure*}

%
\begin{figure}
	\includegraphics[width=\columnwidth]{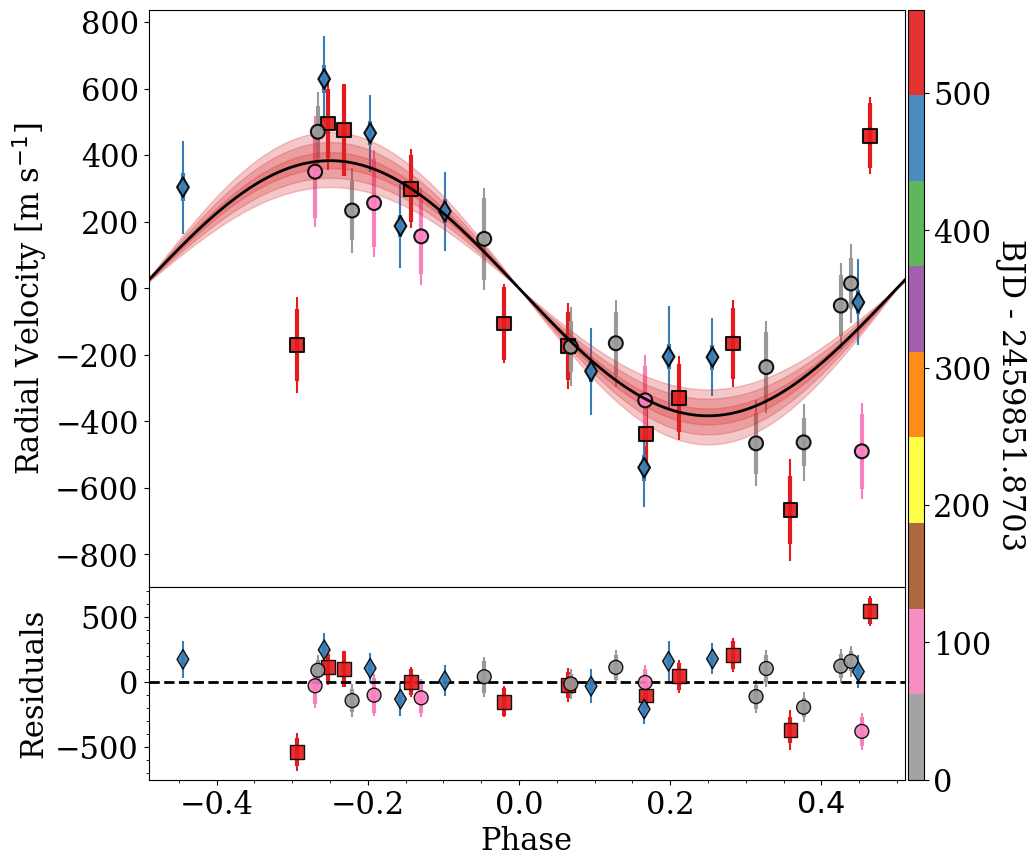}
    \caption{\textbf{Top}: RV phase-folded to the best-fitting period listed in Table 5. RV data is colour-coded in time with the black line and light red shaded region showing the median transit model and its 1, 2 and 3-$\sigma$ confidence intervals. FEROS, HARPS and CORALIE RVs are highlighted by the circles, diamonds and squares, respectively. \textbf{Bottom}: residuals to the best fit model.}
    \label{fig:rvsphot}
\end{figure}
\begin{table}
\centering
\caption{NGTS and TESS photometry for \Nstar. The full Table is available in a machine-readable format from the online journal.  A portion is shown here for guidance.}
\label{tab:ngts}
\begin{tabular}{cccc}
Time	&	Flux       &Flux & Instrument\\
(BJD$_{\rm TDB}$-2457000)	&	(normalised)	&error & \\
\hline
...  &   ...    &   ...  &  ...  \\
1491.65978& 1.0004&  0.0005& TESS\\
1491.68061& 1.0003&  0.0005& TESS\\
1491.70144& 0.9995&  0.0005& TESS\\
1491.72228& 0.9991&  0.0005& TESS\\
1491.74311& 0.9991&  0.0005& TESS\\
...  &   ...    &   ...  &  ...  \\
1836.59516& 0.9964& 0.0019& NGTS\\
1836.59658& 1.0012& 0.0032& NGTS\\
1836.59792& 0.9944& 0.0029& NGTS\\
1836.59933& 0.9969& 0.0014& NGTS\\
1836.60074& 0.9972& 0.0023& NGTS\\
...  &   ...    &   ...  &  ...  \\
2201.74242&  0.9983& 0.0009& TESS\\
2201.74936&  0.9988& 0.0009& TESS\\
2201.75631& 0.9992& 0.0009& TESS\\
2201.76325& 1.0011& 0.0009& TESS\\
2201.77020&  1.0015& 0.0009& TESS\\
    ...  &   ...    &   ...  &  ...  \\
2229.04839&  0.9985& 0.0009& TESS\\
2229.05534& 0.9997& 0.0009& TESS\\
2229.06228&  1.0003& 0.0000& TESS\\
2229.06923& 1.0005& 0.0009& TESS\\
...  &   ...    &   ... & ...  \\
2962.80386 & 0.9977 & 0.0006 & TESS\\ 
2962.80849 & 1.0031 & 0.0018 & TESS\\ 
2962.81544 & 0.9989 & 0.0008 & TESS\\ 
2962.82238 & 0.9995 & 0.0025 & TESS\\ 
2962.82933 & 1.0002 & 0.0012 & TESS\\
\hline
	\end{tabular}
\end{table}

\subsection{TESS Photometry}
\label{sub:tessphot}
TESS \citep{TESS} observed \Nstar\ in Sectors 07, 33, 34 and 61 with cadences of 30, 5, 5, and 3.33 minutes, respectively. The data were acquired from the Mikulski Archive for Space Telescopes (MAST), using the \texttt{lightkurve} package \citep{lightkurve}. The first three sectors were available through distinct data reduction teams, to name a few, the Science Processing Operations Center \citep[TESS-SPOC;][]{jenkins2016tess, caldwell2020tess}, the Quick Look Pipeline \citep[QLP;][]{kunimoto2021quick}, and the Cluster Difference Imaging Photometric Survey \citep[CDIPS;][]{bouma2019cluster}, with the later providing further evidences on the young nature of \Nstar~(see $\S$ \ref{subsub:Agestimation}). Sector 61 is thus far only available through the QLP pipeline, thus upon analysis of the data we opted to use TESS-SPOC lightcurves for Sectors 07, 33, and 34 as the data showed slightly less out of transit dispersion. Fig. \ref{fig:tessphot} shows the detrended phase-folded lightcurve as well as the best-fitting transit model derived in $\S$ \ref{sub:globalmodeling}.

Lastly, we point out that the planet was independently detected by NGTS, despite TESS first observed the star in January 2019 with the release of sector 07. The 30-minute cadence photometry did not provide sufficient evidence for a planet candidate at that time, with a TOI alert issued nearly four years later, on May 3, 2023, by which point NGTS had already gathered evidence for a confirmed planet. NGTS monitoring spanned from March 2019 to June 2021, during which multiple transit events were detected prior to the release of TESS sector 33 in December 2020. Finally, our radial velocity follow-up began on April 14, 2022, and continued until April 12, 2024, with \Nplanet~being labelled as TOI-6442 during this period.

\subsection{Spectroscopic Follow Up}
\label{sub:spect}
\subsubsection{FEROS}
\label{subsub:FEROSspect}
19 high-resolution echelle spectra were obtained for \Nstar~ during UT 2022-04-14 through 2023-01-16 under the FEROS program ID 0110.A-9035(A) (PI: JIV) on the MPG/ESO 2.2-m \citep{kaufer1999commissioning} telescope at the La Silla Observatory in Chile. The 500-1200 sec exposure times spectra were reduced using the automated {\sc ceres} pipeline \citep{brahm2017ceres}, which performs all the steps in the CCF reduction, optimally extracts the spectra and performs wavelength calibration, corrects for instrumental drift, and normalises the continuum. {\sc ceres} computes RVs using the cross-correlation function (CCF) method using a G2 mask. Additionally, we obtain the bisector velocity spans (BIS) which track stellar activity, with any correlation between the BIS and the RVs providing evidence for instrumental, and/or stellar effects impacting the observed spectra. Finally, upon inspection of the reduced data, we dropped the first 4 RVs given their low SNRs and performed the global model in $\$$ \ref{sub:globalmodeling} with 15 FEROS RVs.
\subsubsection{HARPS}
\label{subsub:HARPSspect}
We obtained 10 high-resolution echelle spectra for \Nstar~ during UT 2024-01-06 through 2024-01-10 under the HARPS program ID 112.25QD (PI: DRA) on the ESO 3.6 m \citep{2003Msngr.114...20M} telescope at the La Silla Observatory in Chile. The high accuracy mode (HAM) was used, where we achieved a typical signal-to-noise (SNR) of 35 per pixel at 6500 \AA\, and an RV precision of $\sim37$ \ms\, with exposure times of 1800-2100 seconds depending on weather and seeing conditions. The RV measurements were computed with the standard HARPS pipeline \citep{lovis2007new} using the following binary masks for the cross-correlation:
G2, K5, K0, and M4, where agreement was found amongst the
RVs estimated with these binary masks. Given that \Nstar\, is an F-type star, we adopted the RVs measured from the cross-correlation with a G2 binary mask as it is the closest in spectral type to our target.
\subsubsection{CORALIE}
\label{subsub:CORALIEspect}
The Swiss 1.2m \citep{queloz2000coralie} at the La Silla observatory has been used to collect 11 high-resolution spectra with the CORALIE spectrograph. It has two fibers, with one centred on the target while the other directed to either the sky or the Fabry-P\'{e}rot (FP) etalon for purposes of background subtraction or simultaneous drift calibration, respectively. Our observations made use of the FP mode, where a spectral resolution of $\sim$ 60,000 and wavelength coverage of 390–680 nm allowed us to achieve a typical RV precision of $\sim$ 100 \ms during the data acquisition on 2024 February 27 through 2024 April 12 with 30 min exposure times. The spectra was reduced by the standard CORALIE pipeline, and RVs determined through the cross-correlation method with a binary G2 mask \citep{pepe2002coralie}. Further line diagnosis (e.g., FWHM-CCF and BIS) were also extracted during the data reduction. 
\subsubsection{Spectral diagnosis}
\begin{figure}
	\includegraphics[width=\columnwidth]{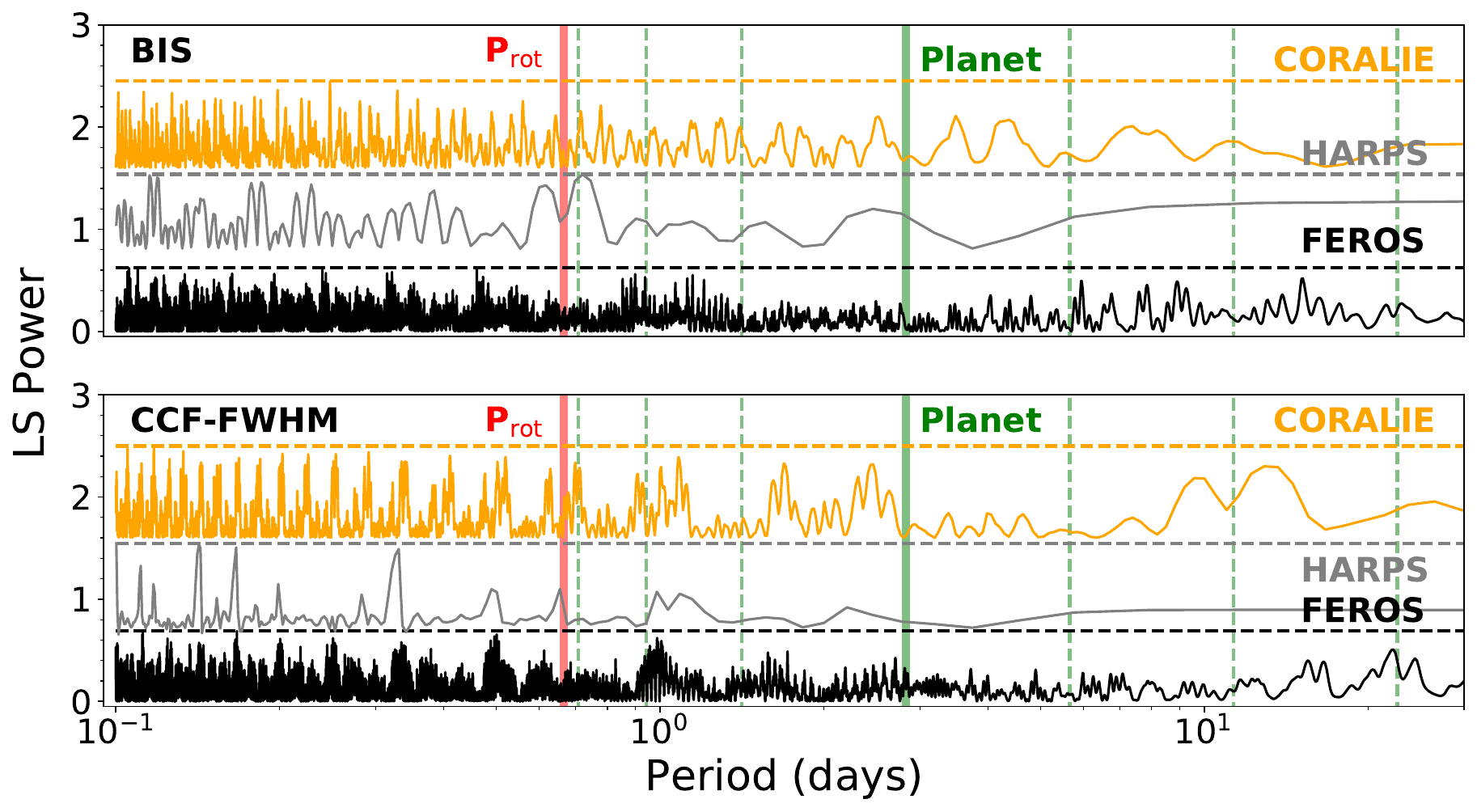}
    \caption{Periodogram of the line Bisectors (top panel) and CCF-FWHM (bottom panel) for FEROS (in black), HARPS (in gray) and CORALIE (in yellow). The false alarm probability level at the highest peak are shown as horizontal black, gray and yellow lines for FEROS, HARPS, and CORALIE, respectively. The BIS and CCF-FWHM FAPs at highest peaks for FEROS, HARPS, and CORALIE are 0.06, 0.59, 0.05 and 0.04, 0.01, 0.02, respectively.
    The planet's orbital period is highlighted by the green vertical line with dashed lines showing the alias 1/8, 1/4, 1/2, 2,3 and 4 from left to right, respectively, and the stellar rotation marked by the red vertical line}.
    \label{fig:diagnosis}
\end{figure}
Radial velocities rely on the accurate measurement of spectral lines centroids. However, stellar activity could affect spectral line profiles, producing spurious periodic RV patterns that could mimic a planetary signal. Diagnosis of spectral lines are a common approach that provides further confidence on the planetary nature of the signal, where correlation between the RVs with parameters such as activity indexes, line bisectors (BIS) or their full width at half maxima (FWHM) are a proxy for activity induced RV variation, rather than a planetary nature. Therefore, we estimate the correlation between RVs against BIS and FWHM-CCF with the Pearson coefficient in Eq. \ref{eq:PearsonEq},
\begin{equation}
\label{eq:PearsonEq}
   r = \frac{\sum(RV_i - \Bar{RV})(x_i -\Bar{x})}{\sqrt{\sum(RV_i - \Bar{RV})^2\sum(x-x_i)^2}} 
\end{equation}
where $x_i$ and $\Bar{x}$ represents the individual data points and their mean value for either BIS or FHWM. The $r$ coefficient between the RVs and BIS for FEROS, HARPS, and CORALIE are 0.193, 0.081, and 0.32, respectively, while for the FWHM-CCF for FEROS, HARPS, and CORALIE, the $r$ values are 0.02, 0.04, and -0.26, respectively. The pearson $r$ coefficient limits are -1 to 1, with values close to zero indicating that little correlation is present in the datasets. Finally, Figure \ref{fig:diagnosis} shows the LS periodogram for BIS and FWHM computed from FEROS, HARPS and CORALIE data, an offset was applied for better visualisation. The false alarm probability (FAP) at the respective instrument highest periodogram peaks are marked as horizontal dashed lines, with no significant peak present at the orbital period of \Nplanet.\\

Lastly, the Gaia DR3 catalogue was examined around \Nstar~to exclude potential contamination from a background eclipsing binary. This search identified five neighbouring stars within the 21-arcsecond TESS pixel, each with G magnitudes between 6.4 and 9.2 fainter than \Nstar, and renormalised unit weight error (RUWE) values below 1.3. Although high-spatial-resolution imaging is unavailable, the NGTS pixel scale is four times finer than TESS, meaning that only one of these stars falls within the NGTS pixel. This star, named GAIA DR3 5590415817155700480, lies 3.1 arcseconds from \Nstar~and has a G-band magnitude of 20, thus contributing negligible flux in the aperture. Its distance also places it outside the 1” HARPS fibre. Additionally, the HARPS CCFs were inspected and show no evidence of wings or secondary dips typically seen in spectroscopic binaries, thus given our dataset, it is unlikely that \Nstar~is part of a binary systems.

\begin{table}
	\centering
	\caption{\Nstar\, follow-up Radial Velocities from FEROS, HARPS and CORALIE}
	\label{tab:rvs}
	\tabcolsep=0.11cm
	\begin{tabular}{cccc} 
BJD$_{\rm TDB}$		&	RV		&RV err & INSTRUMENT\\
-2457000	& (\ms)& (\ms)&(\ms) \\
		\hline

2852.87014602 & 19351.7 &89.7 & FEROS\\
2853.85667801 &18666.5 &94.2 & FEROS\\
... & ... & ... & ... \\
3317.55154657  & 18602.2 &	32.4 & HARPS\\
3318.55134786  & 18553.0 &	36.6 & HARPS\\
... & ... & ... & ... \\
3396.64685773  & 18466.2 &	101.3 & CORALIE\\
3398.57967810  & 19397.1 &	102.4 & CORALIE\\
		\hline
	\end{tabular}
\end{table}
\section{Data Analysis}
\label{sec:analysis}
\subsection{Stellar Rotation Analysis}
\label{sub:rotation}
The star's magnetic fields can prevent local convective motion in the photosphere, consequently blocking the radial heat transfer, thus leading to the appearance of darker and cooler spots compared to the surroundings. Such regions are associated to stellar spots, and are known to be more frequent with increasing stellar activity. As the star spins, its motion brings these spots in and out of sight, causing a periodic brightness variability in the photometric time-series. Therefore, the measurement of the stellar rotation is possible assuming the spot motions are solely due to the stellar spin.

\begin{figure}
    \includegraphics[width=\columnwidth,angle=0]{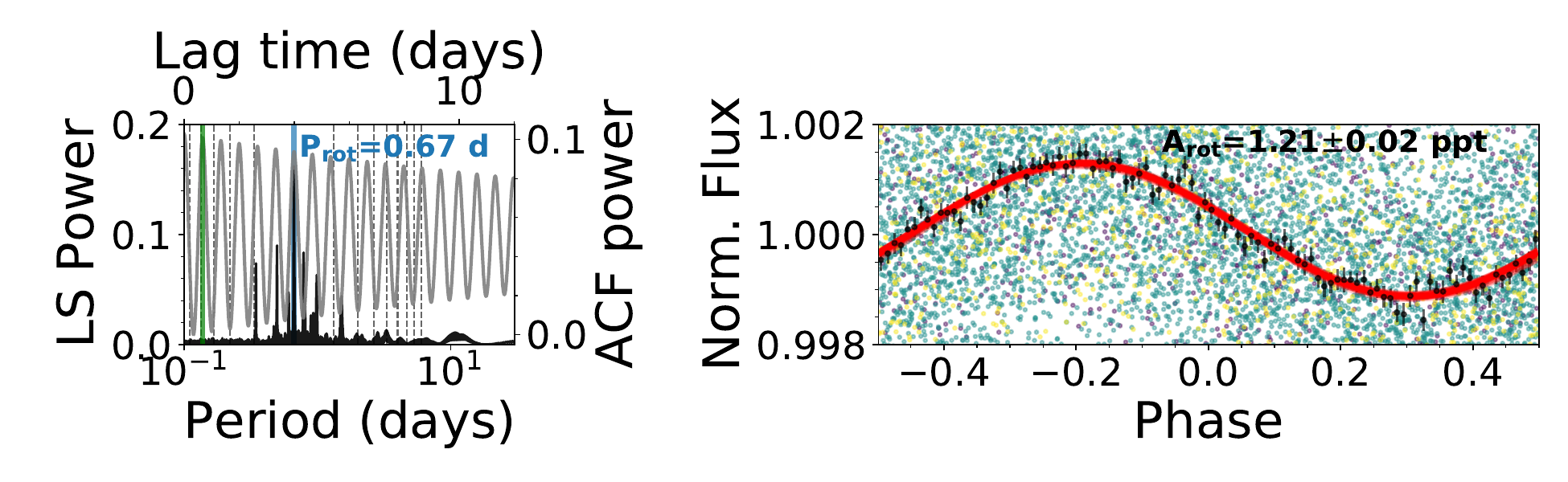}
    \includegraphics[width=\columnwidth,angle=0]{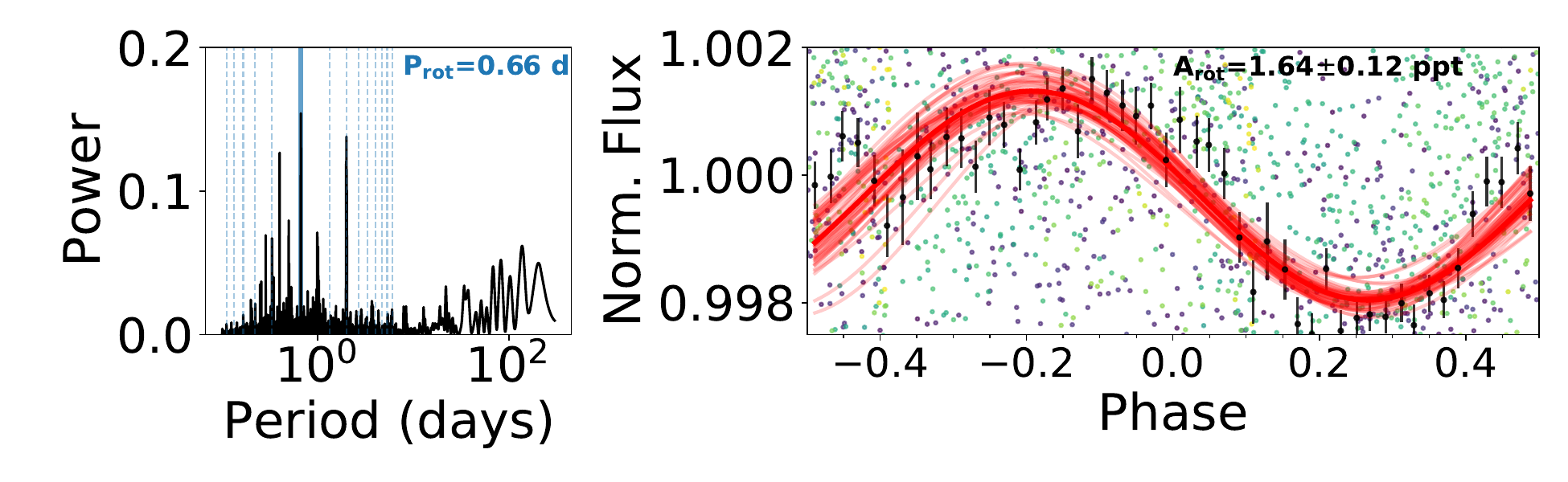}
        \caption{
        \textbf{Top:} The left panel shows the SPOC-TESS LS periodogram in black and ACF in gray, with the bottom and left axes representing the LS periodogram, while the top and left axes are for the ACF model. The vertical blue and green bars corresponding to the optimal periods from LS and ACF, respectively.  The blue coloured P$_{\rm rot}$ at the top right shows both the period at maximum power of the LS periodogram and the period from the ACF. The right panel shows the TESS data folded to the LS best-fitting period colour-coded by time evolving from yellow to purple. Black points are binned in phase domain, with a sinusoidal model and 50 realisations drawn from the final distribution shown in red. A$_{\rm rot}$ shows the median and 1-$\sigma$ best-fitting model.\\
    \textbf{Bottom}: The same as described in the top panel for NGTS data. The ACF is not shown as it did not converged likely due to the gaps in the NGTS data.}
    \label{fig:rotation}
\end{figure}

Following the procedures laid out in previous NGTS rotation works \citep{gillen2020ngts,smith2023ngts}, we derived \Nstar\, rotation period with the Lomb-Scargle \citep[LS;][]{vanderplas2018understanding} and the Auto-Correlation Function \citep[ACF;][]{kreutzer2023s} methods. Both methods rely on distinct assumptions, thus having advantages and limitations. The LS method assumes a sinusoidal function to model the rotation signal, mostly suited for stars that present steady photometric variability, with spot timescales larger than the rotation period. On the other hand, the ACF technique is a more flexible and model-free technique, which measures the degree of similarity between different parts of the dataset \citep[see,][]{mcquillan2014rotation,gillen2020ngts}. To derive the best-fitting period from the ACF we used the undamped simple harmonic oscilator (uSHO) described in Eq \ref{eq:usho}, with $\tau$ defining the decay timescale, $A$ and $B$ to adjust for the ACF power, and $y_0$ to adjust for the offset.
\begin{equation}
\label{eq:usho}
y(t) = e^{\frac{-t}{\tau}}(A\cos{\frac{2\pi t}{P_{\rm rot}}} + B\cos{\frac{4\pi t}{P_{\rm rot}}}) + y_0
\end{equation}
Prior to the period search, we masked the transits, binned the data to 30 minute cadences and median normalised it. Fig \ref{fig:rotation} shows the LS periodogram and the ACF model in the top panel left. The green and blue vertical bars shows that both methods agree to the same 0.67d periodic signal. The top right panel shows the folded lightcurve to the LS period at maximum power. We estimated a P$^{\rm TESS}_{\rm rot}$ = $0.6654 \pm 0.0006$ and A$^{\rm TESS}_{\rm rot}$ = $1.21 \pm 0.02$ ppt from LS periodograms with uncertainties from bootstraping with 10,000 iterations. For the NGTS photometry, the LS periodogram showed the 1-day peak as the main signal, which is commonly associated with the day-night cycles in ground-based missions. Upon removing it by applying a Savitzky–Golay filter, we did the same procedures described above to probe for rotation signatures. Fig \ref{fig:rotation} in the left panel shows the cleaned NGTS data periodogram, where the rotation peak of 0.66 day is present, and followed by its 3$^{\rm rd}$ harmonic marked in vertical dashed lines. The rotation period and amplitude estimated from NGTS photometry is P$^{\rm NGTS}_{\rm rot}$ = $0.665 \pm 0.001$ and A$^{\rm NGTS}_{\rm rot}$ = $1.66 \pm 0.08$ ppt, with uncertainties estimated from bootstraping with 10,000 iterations. The ACF applied to NGTS did not converge due to the several gaps inherent in ground-based missions. The rotation periods from both instruments are in statistical agreement, while the NGTS amplitude is $\sim$ 6-$\sigma$ away from that measured by TESS. Such amplitude mismatch between TESS and NGTS is likely due to spot evolution between measurements from NGTS and TESS, though the instruments distinct photometric band-passes might have also played a role as spots contrast vary with wavelengths or a combination of both.

As TESS data is comprised of 3 distinct sectors, we repeated the analysis above on a per-sector basis. Although we recover the same 0.67 days signal, a second signal at $\sim0.8$ days is present and dominant in sector 7, yet with a LS power difference of $\sim1\%$. Yet, for sectors 33 and 34, the 0.67 days signal becomes more significant with a power difference of $\sim7 \%$ and $\sim10 \%$, respectively, thus why the rotation from the joint lightcurve favoured the 0.67 days period. In addition, the periods derived from ACF for sectors 7, 33, 34 are 0.8, 0.66, and 0.67 days, respectively, thus confirming the LS peaks. We have also carried out the same analysis for QLP lightcurves, and reached the same periods preferences from LS and ACF methods, thus implying that the results are pipeline independent and likely astrophysical. The LS periodogram on the QLP sector 61 showed the 0.8 and 0.67 days periods at similar power, whereas the ACF period was 0.66 days. The NGTS periodogram shows the 0.67 d peak and its harmonics at a much larger power (Fig. \ref{fig:rotation} bottom) compared to the 0.8 day, which after subtracting the 0.67 d, it becomes apparent, with a measured A$_{\rm rot}$ of $0.90 \pm 0.11$, and power difference between the periods of $\sim13\%$., thus weaker than the main 0.67d signal.
Although both 0.67 and 0.8 days signals are significant in the per-sector TESS analysis, we adopted the 0.67 days as it is (1) the dominant signal from the analysis in the joint lightcurve from both LS and ACF methods, (2) it is also the dominant peak in the NGTS data, and (3) its larger amplitude signal compared to the 0.8 d amplitude in TESS ($0.63 \pm 0.03$ ppt) and NGTS ($0.90 \pm 0.11$ ppt). Nonetheless, the 0.67 and 0.8 day signals are likely astrophysical as they are present in both instruments. It is possible that due to the nature of fast rotators, spots could have been propelled towards higher latitudes through the coriolis force. As spots typical lifetimes are of a few days to months, it may be possible that TESS captured distinct spot distributions over sector 7 where the 0.8 day dominates, and the two consecutive sectors 33 and 34, with dominance on the 0.67 d signal. In order to rule out that any of the signals has its origin on UCAC4 271-014742, the nearby star labeled as number 2 in Fig \ref{fig:tesscutfile}, we repeat the rotation analysis for this star using the uncontaminated NGTS photometric time-series. We confirm the rotation period of $\sim$ 0.11 d from GAIA, and derive an amplitude of A$=7.26\pm0.26$ ppt as shown in Figure \ref{fig:UCAC4_contaminant}. As the 0.67d and 0.8d are not present in the LS periodogram, with the ACF method is agreement with the LS period at maximum power, we conclude that both 0.67 and 0.8 days signals are likely originated from \Nstar~rather than UCAC4 271-014742.

We investigated the possibility that the lightcurve periodicity are related to pulsations rather than rotation. \Nstar~ stellar properties place it just outside the bottom of the instability strip near the class of $\gamma$ Doradus. Such stars are late A to early F spectral types, with masses between 1.4 and 2.0 \msun\,, and multi-periodic g-mode induced variability with amplitudes no larger than 0.1 mag and periods of the order of 1 day. Although we do not completely rule out the pulsating nature of these signals, spot-crossing events observed in TESS provide robust evidence for the rotation nature of the photometric periodicity. Fig. \ref{fig:spot-crossing} shows these events, with the first two transits from top to bottom displaying features consistent with spot-crossing, while the last transit, we interpret as a spot entering the line-of-sight, thus reducing the flux. Moreover, spot evolution could well explain the power variability of the 0.67 and 0.8 day signals due to distinct spot distributions as a function time. Additionally, line blending due to rotational broadening evidenced by the lack of Li as well as Fe lines to trace the host age and the projected velocity (v$\sin{i}_{\ast}$), respectively, provides further evidence that the photometric variability are indeed from a spot modulated time-series. Finally, an estimated value of v$\sin{i}_{\ast} \sim$ 111.87 kms$^{-1}$ was computed from the star's radius and adopted rotation period from Table \ref{tab:stellar}. A measurement of v$\sin{i}_{\ast}$ was not possible due to a combination of low SNR, e.g., HARPS SNR 28-35, and line blending. However, even though spectral lines are affected by several physical mechanisms (e.g., collisional broadening) depending on wavelength and stellar properties (e.g., \logg, T$_{\rm eff}$), the CCF shape provides a rough estimate of the averaged spectral line profile, from which we computed the HARPS FHWM-CCF to be $\sim$ 35-38 km/s. Rewriting the v$\sin{i}_{\ast}$ relation as $i_{\ast} = \arcsin{\frac{v\sin{i}_{\ast}}{2\times\pi\rstar/{\rm P}_{\rm rot}}}$ \citep[see e.g.,][]{hirano2014measurements}, and using the FHWM-CCF as proxy for v$\sin{i}_{\ast}$, we computed i$_{\ast}$ to be about 18-20 degrees. This inclination is in agreement to \citep{dong2023hierarchical}, which found that 72 $\pm$ 9$\%$ of the systems with measured sky-projected stellar obliquities have values less than 40 degrees. We point out that a more complete obliquity analysis will be presented in a follow-up paper where the Rossiter-McLaughlin for \Nstar~will be investigated.
\subsection{Stellar Properties}
\label{sub:stellar}

\begin{figure}
	\includegraphics[width=\columnwidth]{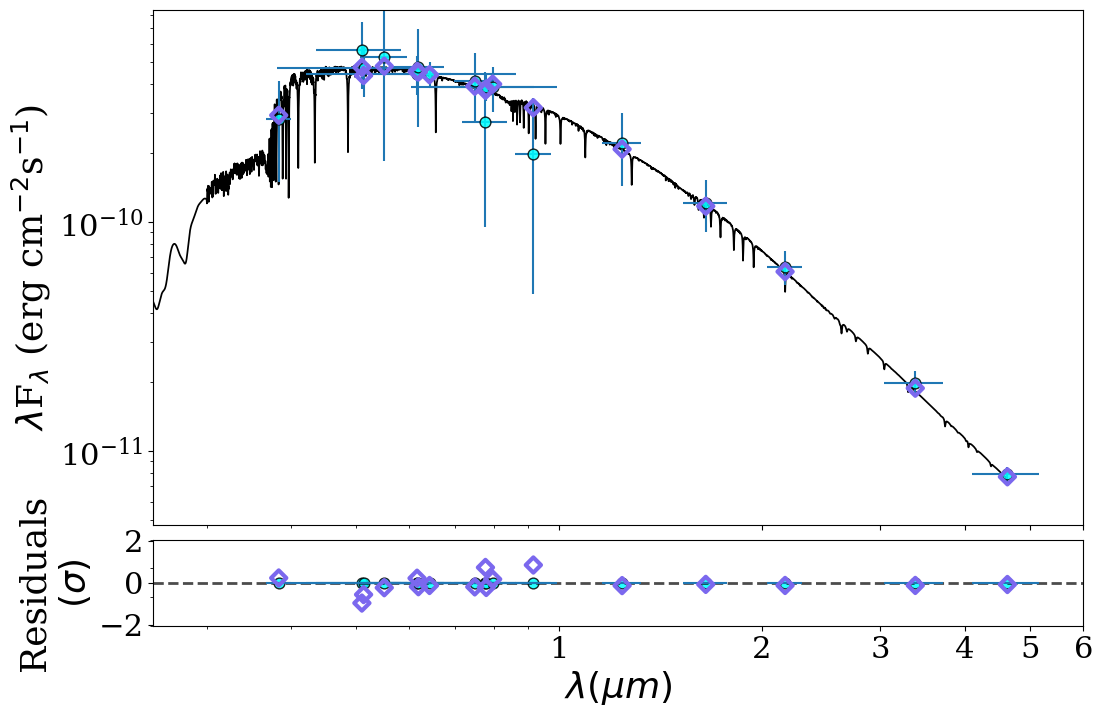}
    \caption{\textbf{Top}: The best-fitting spectral energy distribution model (black line) based on BT-Settl models given the \Nstar\ photometric data (cyan points) and their respective bandwidths shown as horizontal errorbars. Purple diamonds represent the synthetic magnitudes centred at the wavelengths of the photometric data from Table~\ref{tab:stellar}. \textbf{Bottom}: residuals to the best fit in $\sigma$ units.}
    \label{fig:sed}
\end{figure}

\Nstar{} parameters were estimated using the publicly available \texttt{ARIADNE} python package \citep{vines2022ariadne}. The code is based on the spectral energy distribution (SED) fitting method, which consists of fitting archival photometry to synthetic magnitudes from interpolated grids of stellar atmosphere models. The synthetic photometry is computed from convolving a given model with several filter response functions \citep[see available SED models in][]{vines2022ariadne} and scaled by $(\rstar/D)^{2}$, with D being the star's distance to the sun. An excess noise term is introduced for each photometric measurement in order to account for underestimated uncertainties. Finally, a cost function with input parameters T$_{\text{eff}}$, $\log g$, \met, and V band extinction (A$_{\rm V}$) are minimised with \citep[\texttt{DYNESTY;}][]{speagle2020dynesty}, a nested sampling algorithm used to effectively search the parameter space, and find the best set of synthetic fluxes from a given SED model with stellar properties that best matches the observed photometry. \\
\texttt{ARIADNE} performs the above steps for several atmosphere libraries, to name a few, Phoenix V2 \citep{husser2013new}, BT-Settl \citep{hauschildt1999nextgen,allard2012models}, \citet{castelli2004new}, and Kurucz \citet{kurucz1993atlas9}. The code will output stellar posterior distributions from fits using each library, where the adopted stellar parameters are derived from the averaged posterior distributions weighted by their respective Bayesian evidences. Such a Bayesian averaging method helps to mitigate the assumptions and limitations from individual stellar atmosphere model, thus yielding precise stellar parameters, particularly the \rstar~and T$_{\rm eff}$, which are key to inform the global modelling of \Nplanet\ (see $\S$ \ref{sub:globalmodeling}). Finally, T$_{\rm eff}$, \logg, \met~as well as additional quantities such as D, \rstar, and A$_{\rm V}$ from \texttt{ARIADNE} are used to automatically derive the stellar age (\age), \mstar, and the equal evolutionary points from the \texttt{isochrone} package \citep{morton2015isochrones}. \\
We set up \texttt{ARIADNE} with priors defined in Table \ref{tab:priors-ariadne}, where $\mathcal{N}$($\mu$,$\sigma^2$) and $\mathcal{U}$(a,b) define normal and uniform priors with $\mu$, $\sigma^{2}$, a, and b representing the median, variance, and lower and upper limits, respectively. We opted for priors somewhat centred around \Nstar~GAIA DR3\footnote{https://gea.esac.esa.int/archive/} parameters as preliminary tests with broad and uninformative priors were consistent with the values reported by GAIA, which are: T$_{\rm eff}$ = 7213 $\pm$ 9\,K, log g = 4.19$^{+0.03}_{-0.01}$\,dex, \rstar\ = 1.54$^{+0.06}_{-0.03}$\,\rsun\,, distance of 447$^{+17}_{-7}$\,pc, A$_{\rm V}$ = 0.334$^{+0.004}_{-0.001}$, and $\age=0.38^{+0.28}_{-0.18}$ Gyr. We point that the age is computed from the GAIA apsis FLAME module, which is obtained by comparing \Nstar~the T$_{\rm eff}$ and luminosity with the BASTI \citep{hidalgo2018updated} solar metallicity stellar evolution models. \texttt{ARIADNE} computed stellar parameters are: T$_{\rm eff}$ = 7437 $\pm$ 72, log g = 4.26 $\pm$ 0.28\,dex,
\rstar\ = 1.47 $\pm$ 0.06\,\rsun\,, \mstar\ = 1.60 $\pm$ 0.11\,\msun\,, $\rho_{*}$ = $0.71\pm0.12$~\gccc,  distance of 448.4 $\pm$ 7.3\,pc, A$_{\rm V}$ = 0.53$\pm$0.11, and $\age=0.14 \pm 0.12$ Gyr. Fig. \ref{fig:sed} shows the BT-Settl best-fitting SED model to the archival photometry with adopted median and 1$-\sigma$ posteriors distributions for the \Nstar~stellar parameters shown in Table \ref{tab:stellar}.\\

We have also attempted to measure \Nstar~atmospheric properties (T$_{\rm eff}$, \met, \logg, $\sin{\rm i_{\ast}}$) from co-added HARPS spectra using \texttt{SPECIES}\footnote{https://github.com/msotov/SPECIES}~\citep[][]{soto2018spectroscopic}, a code that estimates atmospheric parameters from high-resolution spectra. However, the code could not converge due to a combination of factors, including limited spectral features, unresolved spectral lines for abundance measurement via equivalent widths, and a low signal-to-noise ratio (HARPS-SNR $\sim$ 40), partly due to its faint V$\sim$11.5 mag. As a result, we opted for the SED fitting method instead.

\subsubsection{Age Estimation}
\label{subsub:Agestimation}
A typical method to estimate stellar ages consists of isochrone constructions through the usage of grids of pre-computed stellar evolutionary models described by stellar physical properties (e.g., $T_{\rm eff}$, L$_{\ast}$, \met, so forth) that are interpolated to fit a set of observed stellar parameters. Such evolutionary models are rearranged to tracks of fixed ages, named isochrones, from which stellar ages are estimated. Nonetheless, limitations exist to precisely estimate stellar ages through isochrone fitting, given the complexity and strong non-linearity in the process of finding the solution. Therefore, the usage of independent methods such as Gyrochronology as well as probing lithium abundances have been used to increase the confidence on the age limits. 

We probed gyro-ages from empirical models by \citet{barnes2007ages} (hereafter B07) and \citet{mamajek2008improved} (here after M08) against our measured photometric projected stellar rotation P$_{\rm rot}$ from $\S$ \ref{sub:rotation}. The models provide ages as a function of rotation period and B-V colour, from which \Nstar~is placed between 17 and 41 Myr according to M08 and B07 models, respectively, assuming P$_{\rm rot}$ from Table \ref{tab:stellar}. However, in the case of the 2$^{\rm nd}$ highest peak of $\sim0.8$ day, which is present in some of the \texttt{TESS} sectors (see $\S$ \ref{sub:rotation}), the gyro-ages from M08 and B07 would be about 24 and 58 Myr, respectively.

Membership of a star within a cluster or association can allow for more precise constraints to be placed on its bulk properties such as age and metallicity, through an ensemble analysis of all stars in the same group. Using GAIA DR2 data, \Nstar~was highlighted by \citet{cantat2019expanding} as a likely member (71\% likelihood) of Population V in the young Vela-Puppis region, a cluster of stars with similar kinematics which sit nearby the large Vela OB2 association. Through comparison to PARSEC isochrones (Z=0.019) this population was found to have an age of $\sim$ 20-35 Myr, further supporting the younger age derived from gyrochronology. Such youth is also supported by \Nstar~also being included in the selection of young upper-main sequence stars derived by \cite{2018A&A...620A.172Z}, however a precise age for this star was not given in this work. 

Finally, we searched the \Nstar~coadded spectrum for any signs of lithium lines as its abundance is associated to young stellar ages. Due to Li volatility with temperature, its abundance is depleted quickly in stellar atmospheres within the first hundred million years of the star lifetime, hence the existence of photospheric Li could place place age limits to the star \citep[e.g., see][]{christensen2018ages}. Li lines were probed across the coadded spectra, particularly around the strong Li resonant doublet at 6707.775 \AA\, and 6707.926 \AA, however, possibly due to the fast rotating nature of the host, hence high degree of line blending, no clear evidence of Li was detected at the 17 ppt precision around the doublet. Therefore, although \texttt{ARIADNE}'s age of $0.14\pm 0.12$ Gyr is in agreement to other dating methods, we adopted an age upper limit of $\sim50$ Myr given the evidences from the literature \citep{2018A&A...620A.172Z,cantat2019expanding}, the agreement with \texttt{ARIADNE} age estimate as well as our analysis based on Gyrochronology. A lower age limit of $\sim10$ Myr is based on planet structure models is discussed in $\S$ \ref{sub:RadiusInflation}.

\begin{table}
	\centering
	\caption{Stellar Properties for \Nstar}
	\begin{tabular}{lcc} 
	Property	&	Value		&Source\\
	\hline
    \multicolumn{3}{l}{Astrometric Properties}\\
    R.A.		&	\mbox{$07^{\rmn{h}} 11^{\rmn{m}} 20\fs0004$}			&GAIA\\
	Dec			&	\mbox{$-35\degr 51\arcmin 1\farcs 8648$}			& GAIA	\\
	2MASS I.D.	& J07112000-3551019	&2MASS	\\
	TIC I.D.	& 97921547	&TIC	\\
	GAIA DR3 I.D. & 5590415817155428224	&GAIA	\\
	Parallax (mas) & 2.262 $\pm$ 0.012 &GAIA \\
    $\mu_{{\rm R.A.}}$ (\masy) & -4.936 $\pm$ 0.013 & GAIA \\
	$\mu_{{\rm Dec.}}$ (\masy) & 6.538 $\pm$ 0.014 & GAIA \\
    \\
    \multicolumn{3}{l}{Photometric Properties}\\
	V (mag)		&11.458 $\pm$ 0.021 &APASS\\
	B (mag)		&11.89 $\pm$ 0.028	&APASS\\
	g (mag)		&11.655	$\pm$ 0.014	&APASS\\
	r (mag)		&11.425 $\pm$ 0.012	&APASS\\
	i (mag)		&11.385 $\pm$ 0.011	&APASS\\
    G (mag)		&11.394 $\pm$ 0.003		& GAIA\\
    TESS (mag)	&11.101 $\pm$ 0.006	&TIC\\
    J (mag)		&10.607 $\pm$ 0.023		&2MASS	\\
    H (mag)		&10.475 $\pm$ 0.024	&2MASS	\\
    K (mag)		&10.408 $\pm$ 0.021	&2MASS	\\
    W1 (mag)	&10.366 $\pm$ 0.023	&WISE	\\
    W2 (mag)	&10.381 $\pm$ 0.019	&WISE	\\
    W3 (mag)	&10.508 $\pm$ 0.067	&WISE	\\
    
    \\
    \multicolumn{3}{l}{Derived Properties}\\
    $\rho_{*}$ (\gccc)   & 0.81$\pm$ 0.02        &\texttt{EMPEROR}\\
    $\gamma_{\rm RV-HARPS}$ (\kms) & 18.93 $^{+0.01}_{-0.07}$     &\texttt{EMPEROR}\\
    $\gamma_{\rm RV-CORALIE}$ (\kms) & 18.72 $^{+0.02}_{-0.2}$     &\texttt{EMPEROR}\\
    $\gamma_{\rm RV-FEROS}$ (\kms) & 18.90 $^{+0.07}_{-0.01}$     &\texttt{EMPEROR}\\
    $\sigma_{\rm HARPS}$ (\ms) & 141 $^{+59}_{-13}$     &\texttt{EMPEROR}\\
    $\sigma_{\rm FEROS}$ (\ms) & 90 $\pm$ $^{+39}_{-29}$     &\texttt{EMPEROR}\\
    $\sigma_{\rm CORALIE}$ (\ms) & 299 $\pm$ $^{+86}_{-37}$     &\texttt{EMPEROR}\\
    P$_{\rm rot}$ (days) &  0.6654 $\pm$ 0.0006 & This work\\

    T$_{\rm eff}$ (K)    & 7437 $\pm$ 72       &\texttt{ARIADNE}\\
    $[$Fe/H$]$	 & -0.11 $\pm$ 0.17       &\texttt{ARIADNE}\\
    log g                &	4.26$\pm$ 0.28&\texttt{ARIADNE}\\
    Age	(Myr)		     & 10-50 &\texttt{This work}\\
    \mstar (\msun)       & 1.60$\pm$ 0.11          &\texttt{ARIADNE}\\
    \rstar (\rsun)       & 1.47$\pm$ 0.06 	      &\texttt{ARIADNE}\\
    Luminosity (L$_\odot$)	     & 5.91 $\pm$ 0.54   &\texttt{ARIADNE} \\
    Distance (pc)	     & 438.5 $\pm$ 7.1   &\texttt{ARIADNE} \\
	\hline
    \multicolumn{3}{l}{2MASS \citep{2MASS}; TIC v8 \citep{stassun2018tess};}\\
    \multicolumn{3}{l}{APASS \citep{APASS}; WISE \citep{WISE};}\\
    \multicolumn{3}{l}{{\em Gaia} \citep{brown2021gaia}}\\
	\end{tabular}
    \label{tab:stellar}
\end{table}
\begin{figure}
	\includegraphics[width=\columnwidth]{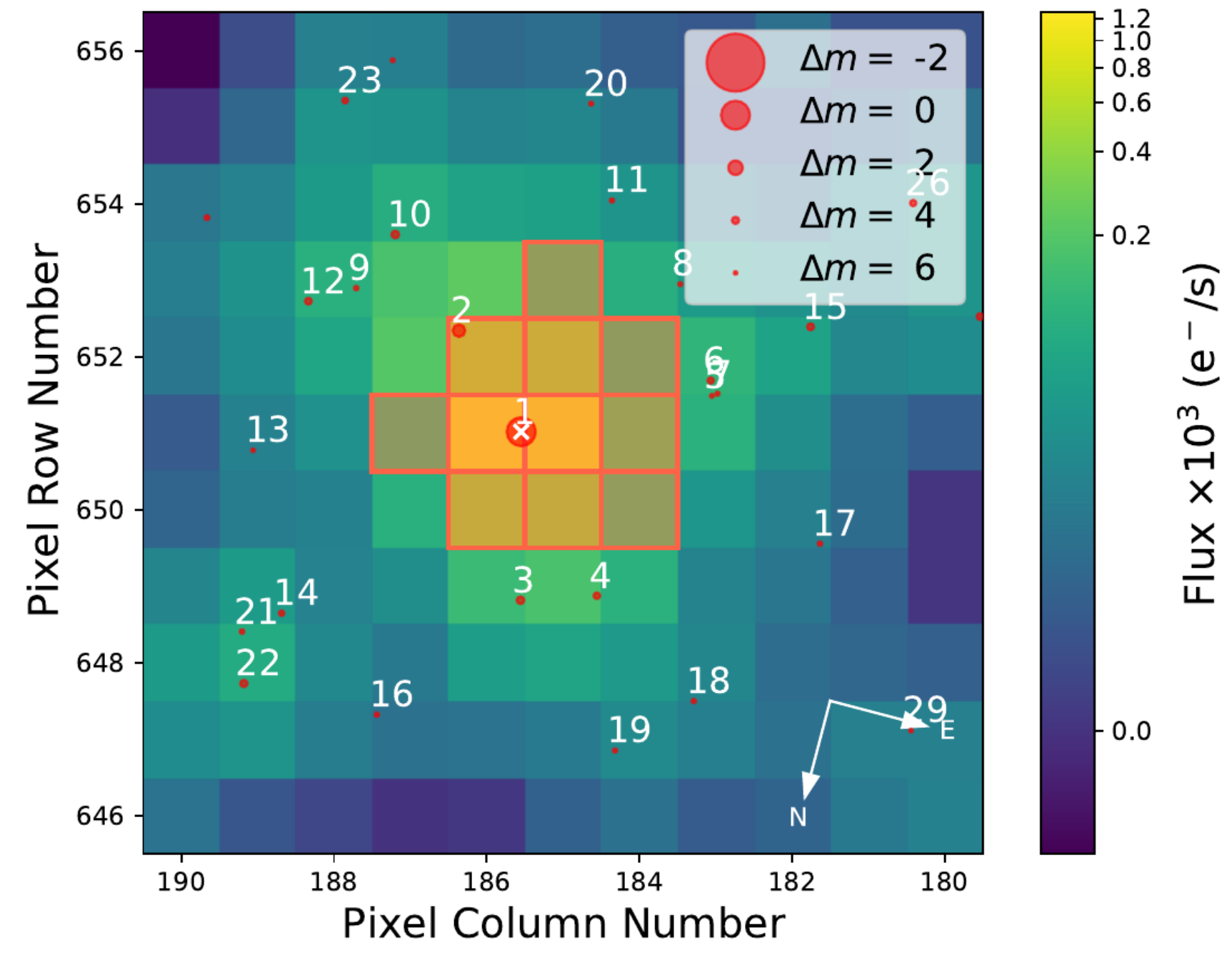}
    \caption{
    TESS Sector 7 Full-Frame Image cutout (11 x 11 pixels) generated with the \texttt{tpfplotter} script described in \citet{aller2020planetary}. \Nplanet\ is shown in the centre labeled number 1, followed by UCAC4 271-014742 (number 2), a V = 13.9 mag, and 32" away from the planet. Our analysis has shown that the star did not contribute significant flux to the aperture, thus negligible dilution was observed in TESS.
    }
    \label{fig:tesscutfile}
\end{figure}
\subsection{Global modelling}
\label{sub:globalmodeling}
We performed the global modelling with the \texttt{EMPEROR} code\footnote{https://github.com/ReddTea/astroEMPEROR} \citep[Pena \& Jenkins in prep]{vines2023dense}, a python toolkit that allows for a joint radial-velocity and photometric analysis. \texttt{EMPEROR} is equipped with Gaussian Processes as well as auto-regressive integrated moving average (ARIMA) models that can be added to the cost function as part of the noise model to properly account for both instrumental and/or astrophysical noise. Moreover, \texttt{EMPEROR} flexibility allows the user to choose amongst samplers such as the Markov-Chain Monte Carlo \citep[\texttt{MCMC};][]{foremanmackey13}, Parallel Tempering MCMC (PTMCMC), and nested sampling \citep[\texttt{DYNESTY;}][]{speagle2020dynesty} to effectively explore the n dimensional parameter space, and reach the optimal solution.

Here we lay out the steps leading to the adopted solution shown in Table \ref{tab:stellar} and Table \ref{tab:planet}. We started by running the code with photometry and RV time-series separately, and found a period difference between the datasets of only $\sim8$ min. For the global model, multiples runs were executed to probe the prior distributions, the inclusion of an ARIMA model or Gaussian Processes as part of the noise model, linear or quadratic acceleration terms. The adopted solution has been selected upon comparison of every run logarithm Bayesian evidence ($\log Z$) as well as their Bayesian Information Criteria (BIC). Both metrics are frequently used in model selection, where higher $\log Z$ and lower BIC values indicate a preference for a particular model. 

As both photometry and RV measurements are commonly impacted by (non-) correlated noise related to the instruments (e.g., nightly drifts, atmospheric turbulence, so forth) and/or the star (e.g., convective overflows, granulation, flares, spots), we found that the runs with GPs included in the photometry time-series yielded low BICs and high $\log Z$ compared to the runs without GPs. In fact, \Nstar~photometry presented clear signs of periodically modulated activity (see $\S$ \ref{sub:rotation}) near 0.67d, thus it was necessary to include a global GP Matern kernel to properly model the spot modulated lightcurve. We have also allowed supersampling on the transit model with 20 points per bin for the TESS sector 7 to account for its 30 minute cadence. The Generalised Lomb-Scargle (GLS) periodogram applied to the RV time-series did not show any significant peak at the rotation period, but a highest peak at the planet period 1.83 days (FAP = $\sim$ 0.1$\%$) followed by the 1.39 day peak (FAP = $\sim$ 8$\%$), with latter closer to the first harmonic of the rotation period. Although we are convinced the signal is related to stellar activity, we searched the lightcurves for any evidence of companions and also computed \Nplanet~transit timing variations, where both analysis were against further companions in the system. Finally, we tested the dynamical stability (in $\S$ \ref{sub:dynamicalStability}) assuming a planetary nature for the 1.39d signal, with results consistent with an unstable system, thus we conclude this signal is indeed related to activity, which was modelled with an additional Keplerian function. 

Finally, we point out that no dilution terms have been included in the photometry. The brightest (V = 13.9 mag) nearby object, UCAC4 271-014742 is 32" away and did not affect the transits in the NGTS time-series given its small pixel scale of $\sim$ 5"/pixel. The TESS full frame images (FFI) though, with an on-sky size of $\sim$ 21", would place the contaminant at $\sim$ 1.5 pixels away, leading to a theoretical upper limit flux dilution of 8.75$\%$ (Fig. \ref{fig:tesscutfile}), had the contaminant been in the same pixel. However, NGTS and TESS transits have the same depth, thus we conclude that the transits from the TESS mission are not severely affected by dilution. As part of the dilution analysis, we checked our target and the contaminant time-series from NGTS photometry, and ascertain that the transits are on-target. Fig \ref{fig:ngtsphot} and \ref{fig:tessphot} show the NGTS and TESS detrended lightcurves with their best-fitting models. The RVs are shown in Fig. \ref{fig:rvsphot}. The parameters of the adopted best-fitting transit and Keplerian models can be found in Table \ref{tab:planet}, while the second Keplerian associated to stellar activity is in Fig. \ref{fig:rvsKep2} with parameters shown in Table \ref{tab:Kep2}.

We note that two CORALIE RVs highlighted in red squares near phases -0.3 and +0.45 located below and above the curve, respectively, disagree significantly with the model. We investigated their BIS, FWHM-CCF as well as the H$\alpha$, Na, Ca line activity indicators and found no evidence that the RVs were affected by stellar activity, thus we kept the points while performing the Global modelling.

\begin{table}
	\centering
	\caption{Planetary Properties for \Nplanet}
	\begin{tabular}{lc} 
	Property	& 	Value \\
	\hline
    P (days)		        & 2.827972 $\pm$ 0.000001	\\
	T$_C$ (BJD$_{\rm TDB}$)		& 2459986.4090 $\pm$ 0.0003	 \\
    T$_{14}$ (hours) & 2.62 $\pm$ 0.02 \\
    $a/\rstar$   & 6.99$\pm$0.07 \\
    \rpl/\rstar & 0.1146 $\pm$ 0.0004 \\
    $b$ & 0.74 $\pm$ 0.01                       \\
    $i(deg)$ & 83.94$\pm$0.12                      \\

	K (\ms) 	&383$^{+25}_{-23}$	                           \\
    e 			& 0.0 (fixed)  	\\
    $\omega~(\deg)$ & 90 (fixed) \\
    \mpl (\mjup)& 3.63$\pm$0.27	\\
    \rpl (\rjup)& 1.64$\pm$0.07  \\
    $\rho_{p}$ (\gccc) & 0.19$\pm$0.03 \\
    a (AU) & 0.048$\pm$0.002 \\
    T$^{\ast}_{eq}$ (K) & 1991 $\pm$ 21 \\
    Flux (\ergscm) & (3.56$\pm$1.53)$\times$10$^9$	\\
	\hline
    \multicolumn{2}{l}{$\ast$ Assumed zero Bond albedo}\\
	\end{tabular}
    \label{tab:planet}
\end{table}

\subsection{Transit Timing Variation}
\label{sub:TTVs}
Transit timing variations \citep*[TTVs;][]{agol2005detecting} occur when mid-transit times $T_0$ deviate from a linear ephemeris model given by $T_n = T_0 + N \times P$, with $N$ and $P$ being the transit number and orbital period, respectively. The most commonly reported scenarios in the literature are: (1) dynamical interactions between the star and the planet, which often results in orbital angular momentum exchanges, leading to orbital decay, and consequently the planet in-falls towards its host star \citep[e.g., WASP-12][]{wong2022tess}. (2) Planet-planet interactions, specifically near mean motion resonance (MMRs) sites, where the interaction becomes more significant, leading to higher TTV amplitudes \citep[e.g., WASP-47][]{becker2015wasp}. In addition, the interacting bodies frequently show anti-correlated TTVs and transit duration variations. In fact, several multi-planet systems in MMRs from the Kepler mission were validated as bonafide planets through the TTV method \citep{cochran2011kepler,gillon2017seven,steffen2012transit}, as the host stars brightness were too dim to allow for the necessary RV precision for mass measurements. However, a few systems had their masses estimated from RVs \citep[][]{barros2014sophie, almenara2018sophie}, while others had their masses and eccentricities measured given the precise TTVs data \citep{lithwick2012extracting}, which revealed the chopping effect, responsible for breaking the M-$e$ degeneracy. Finally, TTVs coupled with RV observations support that hot Jupiters are not part of multi-planet systems \citep{steffen2012kepler, holczer2016transit}, with such results placing major constraints on giant planets orbital evolution.

\begin{figure}
	\includegraphics[width=\columnwidth]{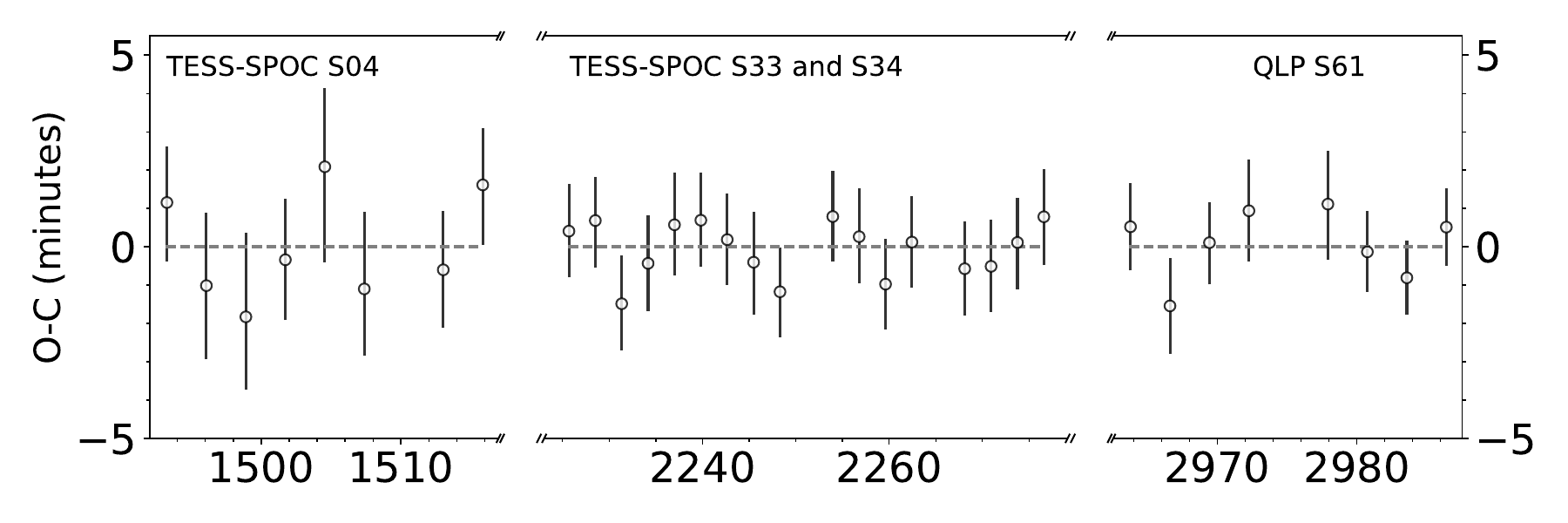}
        \includegraphics[width=\columnwidth]{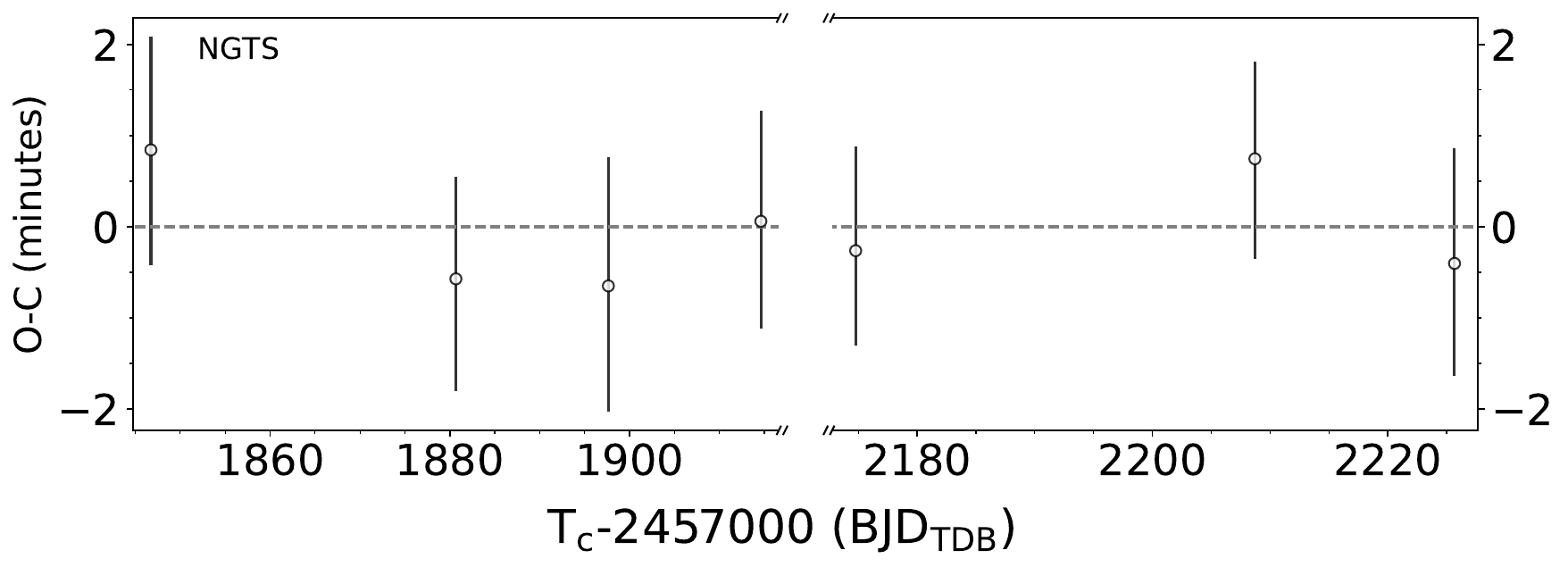}
    \caption{\Nstar\, transit timing variation for the TESS mission (top panel; black open circles) and NGTS mission (bottom panel). The abscissa was zoomed for better visualisation and avoid the large gaps in the time domain. In the top panel, each portion represents a TESS sector with its corresponding reduced pipeline and sector.}
    \label{fig:ttvs}
\end{figure}

\Nplanet~ TTVs were estimated from the GP detrended TESS and NGTS time-series using only transits with full coverage. Each transit mid-time $T_n$ was modelled with the \texttt{batman} \citep{kreidberg2015batman} code and final distributions computed with the affine invariant MCMC code implemented in the \texttt{emcee} package \citep{foremanmackey13}. The transit depth $p = \frac{R_b}{R_*}$, normalised semi-major axis $a = \frac{a}{R_*}$ and a linear offset $bl$ around the normalised flux were free parameters in the model. $T_n$, $p$, and $a$ were assigned the following uniform priors $\mathcal{U}$($T_n$-0.1, $T_n$+0.1), $\mathcal{U}$($p$-0.05, $T_n$+0.05), $\mathcal{U}$($a$-0.5, $a$+0.5), and $\mathcal{U}$($bl$-0.1, $bl$+0.1), respectively, with all other parameters being fixed to their medians from Table \ref{tab:planet}. Finally, a linear ephemeris model was fit to the $T_n$, where we used 10,000 MCMC steps, with 20$\%$ discarded as burn-in. The best-fitting linear model parameters $T_0$ and $P$ for TESS are given by $T_0 = 1493.2406 \pm 0.0003$ days and $P = 2.82796937 \pm 0.00000089$ days, while for NGTS $T_0 = 1838.2528 \pm 0.0005$ days and $P = 2.8279711 \pm 0.0000049$ days. The OC residuals to the linear ephemeris model for both instruments are shown in Fig. \ref{fig:ttvs}, where no evidence for significant TTVs were found, and data made available in table \ref{tab:ttvs} . 
\subsection{\Nstar~ DYNAMICAL STABILITY}
\label{sub:dynamicalStability}
Our global model analysis ($\S$ \ref{sub:globalmodeling}) revealed a planet signal at 2.83 days and rotation signals at 0.67 days and 0.8 days from the photometric time-series. The RV dataset shows the planet signal followed by an additional peak at 1.39 days, which is twice the photometric rotation period, thus we attributed to stellar activity. In order to provide further credibility that the 1.39 days signal is activity related, we performed dynamical stability simulations to initially (1) assure our solution is dynamically stable and also (2) reject the planetary nature of the 1.39 days signal detected in the RV periodogram. For this purpose, we used \texttt{REBOUND}\footnote{https://rebound.readthedocs.io/en/latest/} \citep{rebound}, an N-body problem solver that integrates particles' motion in time under the influence of larger bodies gravitational potential.

Anchored on our orbital solution laid out in $\S$ \ref{sub:globalmodeling} and shown in Table \ref{tab:stellar}, we ran the code up to $\sim$ 100 Myr assuming a 2-body problem made up of our host star and \Nplanet~in order to test the systems stability given our solution. We found no significant changes in its orbital parameters up to this time, otherwise, a larger integration time would have been carried out. Therefore, we stopped the run and assumed the orbital solution is dynamically stable up to $\sim$ 100 Myr. \\ 

\begin{figure}
	\includegraphics[width=\columnwidth]{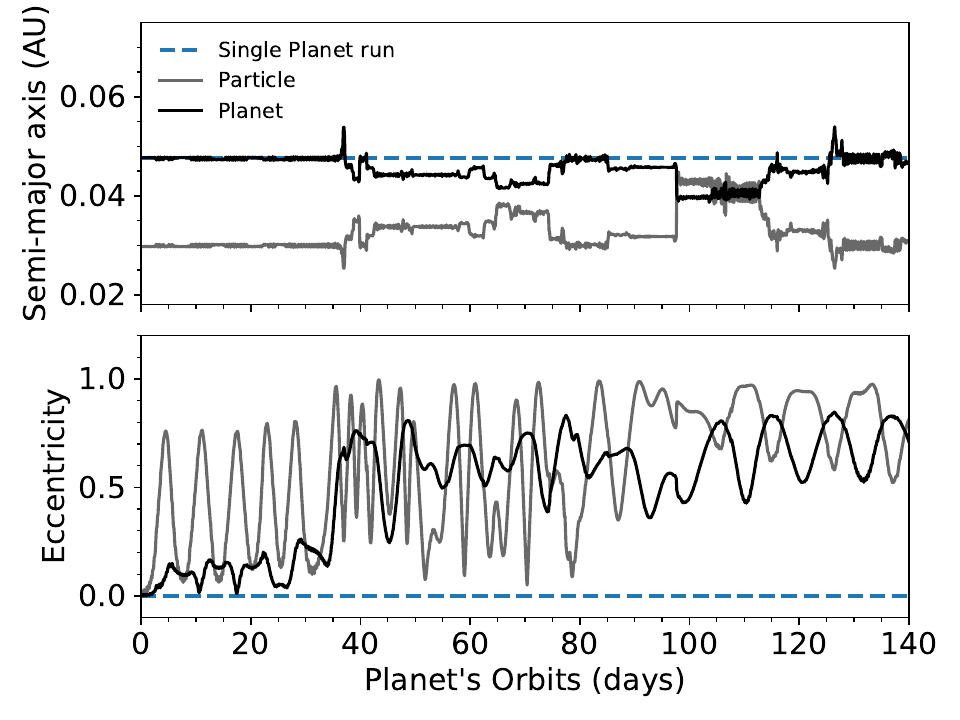}
    \caption{\textbf{Top}: Semi-major axis evolution for \Nstar~ with two planets at orbital periods consistent with \Nplanet~ (in black) and a particle similar to a Jupiter planet with \textbf{$m\sin{i} = 1.29 \mjup$} at 1.39 days (in gray). The dashed blue line corresponds to the orbital evolution of the semi-major axis for when \Nplanet~ is alone in the system. \textbf{Bottom:} Eccentricities evolution. Colours and labels are the same as shown in the top panel.}
    \label{fig:orbEvol}
\end{figure}

In the second scenario containing the host star, \Nplanet, and an inner test particle consistent with the RV signal, i.e., orbital period of 1.39 days and 1.29 $\mjup$. Fig. \ref{fig:orbEvol} showed that the semi-major axes and eccentricities were dynamically unstable, with the particle and planet reaching an eccentricity of 1.0 after nearly 140 orbits of \Nplanet, thus the integration had been stopped due to the chaotic motion. As the particle inclination is unknown, we ran the code multiple times for a 60$^{\circ}$-85$^{\circ}$ inclination grid with 5$^{\circ}$ steps. Every run resulted in the system being unstable. For simulations with nearly co-planar orbits, the planet would have likely been detected during our searches for further planets in the photometric time-series. We point out that chaotic simulations are highly sensitive to the initial conditions, i.e., a change in the particle's parameters result in a slight distinct orbital solution, yet every instance yielded unstable parameters, with the particle frequently being ejected at some point in time. Such chaotic behaviour is expected as both \Nplanet~and the particle are HJs near MMRs. Finally, we conclude that our solution described in the global model is dynamically stable, with the second Keplerian being interpreted as stellar activity, rather than a HJ near an MMR state, which is corroborated by the non TTV detection in $\S$ \ref{sub:TTVs}.
\section{Discussion}
\label{sec:discussion}
\subsection{The transiting hot Jupiter population}
\label{sub:THJ-population}
\begin{figure}
	\includegraphics[width=\columnwidth,angle=0]{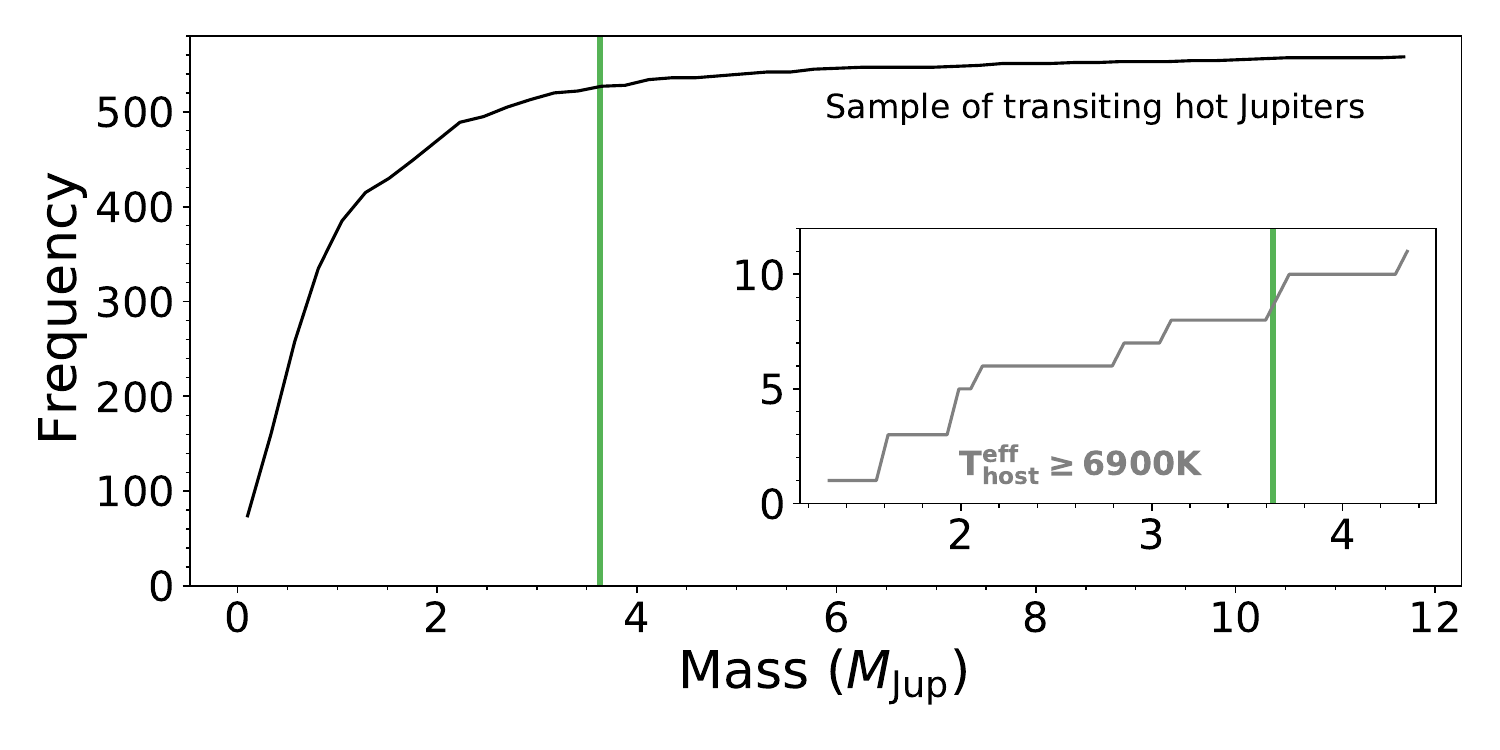}
	\includegraphics[width=\columnwidth,angle=0]{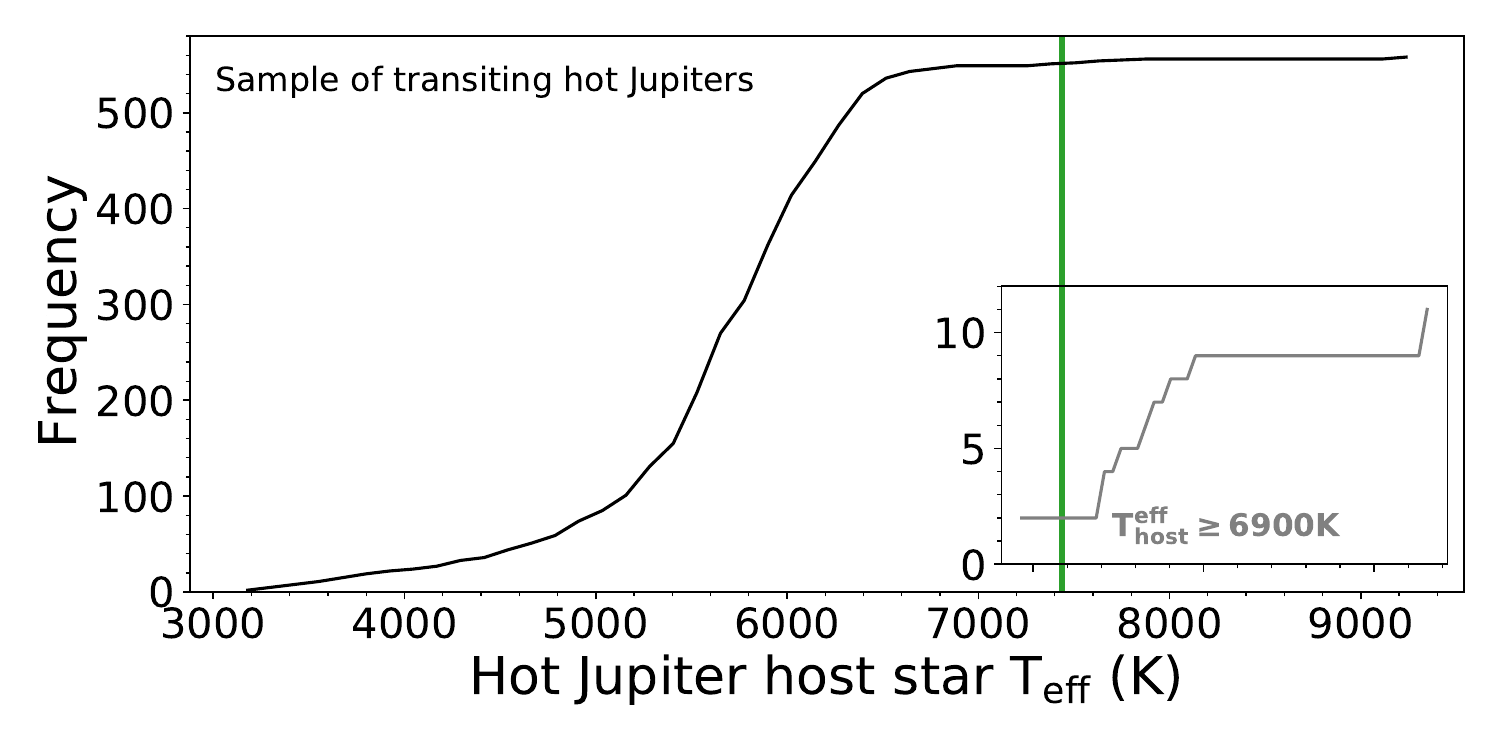}
    \caption{\textbf{Top}: Cumulative distribution functions for the sample of transiting HJs shown as a black solid curve. The green vertical bar highlights \Nplanet's mass. The bottom-right plot shows a slice of the cumulative distribution (in gray) containing HJs whose host effective temperatures are above 6900 K. \Nplanet~ is not included in either plot.
    \textbf{Bottom}: Same as above for the effective temperatures of transiting HJ hosts.  
    }
    \label{fig:HJdemography}
\end{figure}
\begin{figure}
	\includegraphics[width=\columnwidth,angle=0]{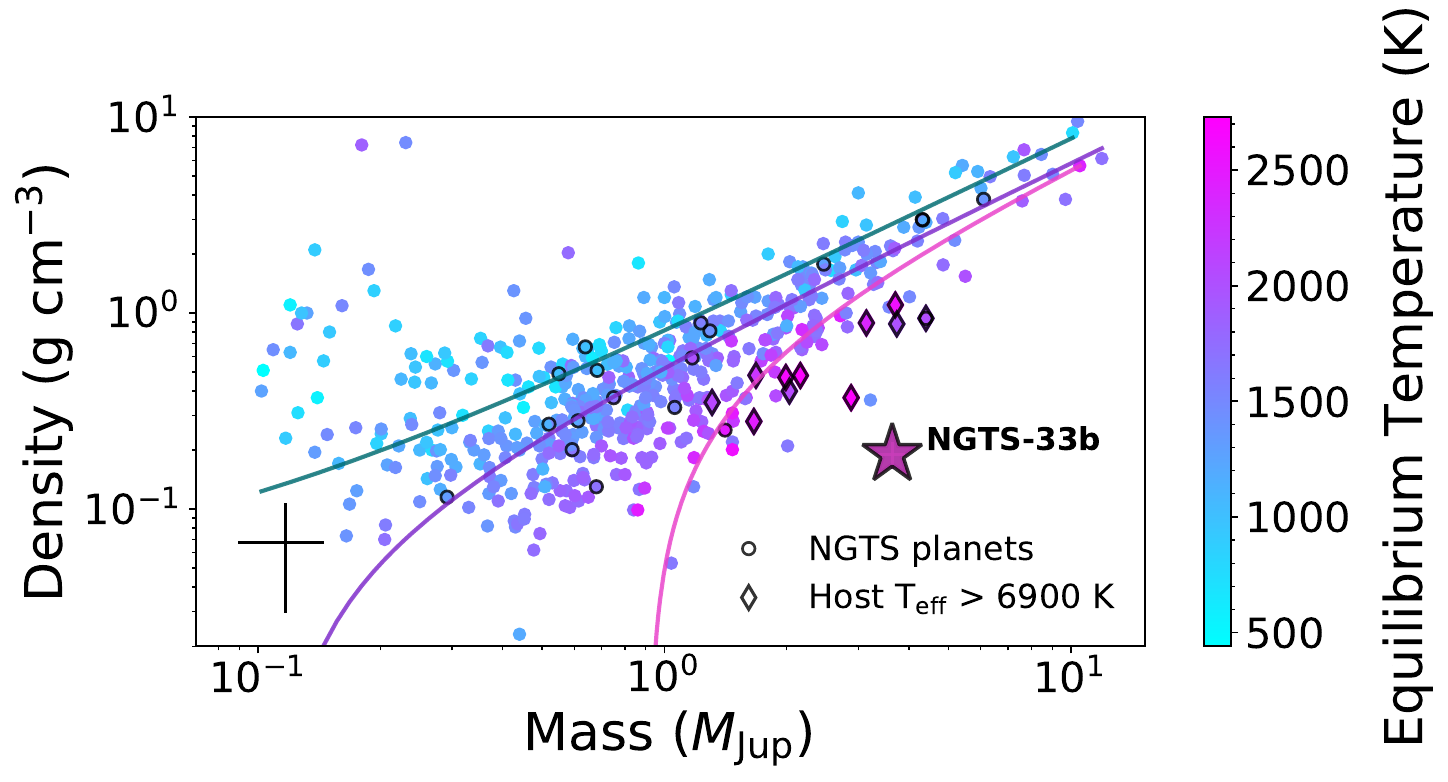}
    \caption{THJ bulk densities as a function of their masses, colour coded by equilibrium temperature. \Nplanet~is represented by the large red star near the image centre towards the bottom right, while the typical uncertainties are shown as a black cross in the bottom left. Black edged diamonds and circles highlight the THJs whose host Teffs are greater than 6900 K, and the population of NGTS detected planets, respectively. Best-fitting empirical linear models describing the THJ populations with distinct T$_{\rm eq}$ are highlighted by the cyan, purple and pink colours. See $\S$ \ref{sub:THJ-population} for explanation.}
    \label{fig:rhoVsMassVsTeq}
\end{figure}
Here we place \Nstar~planetary system in context with the THJ population drawn from the TEPCat catalogue \citep[][]{southworth2011homogeneous}. We selected the planets by P $\leq$ 10 days and mass range 0.1$\mjup$ $\leq$ M $\leq$ 13$\mjup$, yielding 558 giant planets\footnote{We do not include HATS-70 in the analysis as it is at the boundary between planets and brown dwarfs, with a mass upper limit at 1-$\sigma$ of 13.02 $\mjup$} on 2 May 2024. Our analysis laid out in $\S$ \ref{sec:analysis} reveals that \Nstar~is not only the first NGTS discovery of a super-Jupiter around a young fast-rotating host, but also the lowest in density around hot stars (T$_{\rm eff} \geq 6900$K), thus making it an interesting target for transmission and emission spectroscopy follow-up with the James Web Space Telescope (JWST; see $\S$ \ref{sub:TSM-ESM}). 

Fig. \ref{fig:HJdemography} top panel shows the cumulative mass distribution function for the THJ sample, with \Nplanet~belonging to the $\sim$ 7$\%$ most massive THJs discoveries thus far. If we compare it to the scarce population of $\sim$ 11 massive THJ around hot hosts, shown in the embedded plot at the lower right, \Nplanet~is amongst the 2$^{\rm nd}$ most massive THJ detected, alongside planets such as MASCARA-1 \citep{talens2017mascara}, HAT-P-69 \citep{zhou2019two} and OGLE-TR-L9 \citep{snellen2009ogle2}, with masses of M$_{\rm p}$ = 3.7 $\pm$ 0.1 $\mjup$, M$_{\rm p}$ = 3.6 $\pm$ 0.6 $\mjup$ and M$_{\rm p}$ = 4.5 $\pm$ 1.5 $\mjup$, respectively. We note that \Nplanet~mass is in 1-$\sigma$ agreement to them, thus placing it amongst the most massive planets around hot hosts. On the other hand, the lower panel of Fig. \ref{fig:HJdemography} shows the hosts T$_{\rm eff}$ cumulative distribution function. The rapid growth of the slope centred around G dwarfs and spread over K-type ($\sim$ 4000-5250 K) through late F-type stars ($\sim$ 6300 K) represents the majority of THJ hosts. \Nstar~is highlighted by the green vertical bar, and is amongst the $\sim$ 2$\%$ most massive planet hosts, thus reinforcing the rarity of \Nplanet~detection.

In Fig. \ref{fig:rhoVsMassVsTeq} we compared \Nplanet's bulk density with the THJ population, highlighting other NGTS discoveries as well as giants around as hot hosts (T$_{\rm eff} \geq 6900$ K; shown as diamonds in the figure). The $\rho_{\rm p}$ median of similar planets (2\mjup$\leq$M$\leq$5\mjup) is $\sim$ 1.48 \gccc, thus placing \Nplanet's density $\sim$ 13$\%$ below the median, or $\sim$ 22$\%$ less dense, if compared to the $\rho_{\rm p}$ median (0.88 \gccc) of THJ hosted by stars with T$_{\rm eff} \geq 6900$ K. Such discrepancies may be attributed to \Nplanet~ not having undergone significant gravitational collapse due to its young age, thus evidenced by its large radius. However, an irradiated atmosphere, tidal heating during its orbital circularisation or a combination of both could lead to an increase in its radius (see $\S$ \ref{sub:RadiusInflation}), thus significantly reducing its bulk density compared to giants with similar mass.

Finally, in the same figure, distinct THJ groups are observed, which can be distinguished by their T$_{\rm eq}$ and evidenced by the colour gradient. The THJs with higher T$_{\rm eq}$ have significantly lower $\rho_{\rm p}$ compared to cooler planets, thus providing evidence that their bulk densities are a function of T$_{\rm eq}$ regardless of their thermal evolution processes (e.g., high stellar incidences, tidal heating, so forth). For all planetary masses, the T$_{\rm eq}$ correlates positively with radius, even though more massive planets should be less affected by radius anomaly. This is evidenced by the increasing bulk density with lower T$_{\rm eq}$ regardless of mass bin. A thorough analysis of the T$_{\rm eq}$, M$_{\rm p}$ vs $\rho_{\rm p}$ parameter space may even serve as an independent means to estimate planetary radius for non-transiting planets, where their masses, temperatures could be used to estimate a bulk density, and subsequently their radius. A complete analysis of this problem is beyond the scope of this work, however, we performed a linear fit assuming three THJ populations defined by T$_{\rm eq}$ bins. The lower and upper boundaries were initially motivated by visual inspection, but corroborated by several works showing that the radius anomaly is observed roughly at 1000-1400 K \citep[e.g.,][]{demory11,thorngren2018bayesian}.

Therefore, we adjusted linear empirical models in the form of $\rho_{\rm p}$ = A$\times$M + B in order to probe the relationship between $\rho_{\rm p}$ and M to the three populations defined by T$_{\rm eq}<1400$ K (cyan), T$_{\rm eq}\geq1400$ K and T$_{\rm eq}<2300$ K (purple), and T$_{\rm eq}\geq2300$ K (pink), respectively. A and B are free parameters to the model, with best-fitting values given by $0.7706\pm0.0004 ~\mjup^{-1}$ and $0.045\pm0.001$ \gccc, $0.586\pm0.001 ~\mjup^{-1}$ and $-0.065\pm0.004$  \gccc, and $0.5858\pm0.0001~\mjup^{-1}$ and $-0.537\pm0.002$ \gccc, respectively, for the three curves defined above by their T$_{\rm eq}$. The cyan and purple models indicate that giants with lower T$_{\rm eq}$ have larger $\rho_{\rm p}$ regardless of mass, with a typical $\rho_{\rm p}$ differences of 0.18, 0.23 and 0.62 \gccc for giants with masses of 0.6, 1 and 10 \mjup~planets, respectively. Finally, we note that \Nplanet~was not taken into account during the model fitting as its density does not follow any of the three populations.
\subsection{Radius Inflation}
\label{sub:RadiusInflation}
\begin{figure}
	\includegraphics[width=\columnwidth,angle=0]{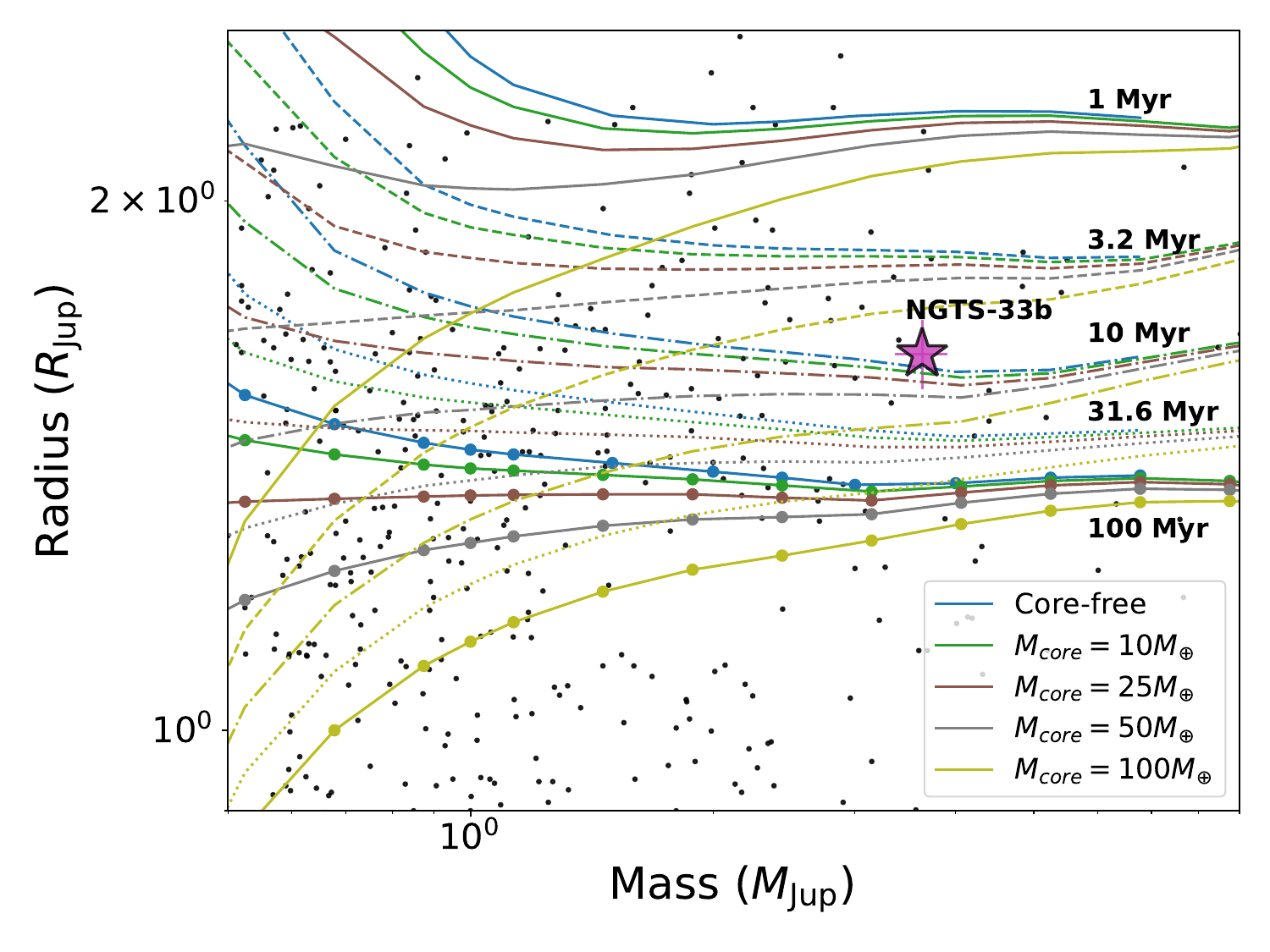}
    \caption{Planet structure models by \citet{fortney2007planetary}. Colour lines represent the same assumed planet core mass, while the lines shapes describes a HJ evolved to a given epoch shown on the right. For instance, the solid lines on the top represent a HJ at 1 Myr, where blue and green lines indicating a core-free and 10M$_{\oplus}$ core mass models. \Nplanet~is highlighted by the purple star symbol and black dots are THJs from the TEPCat catalogue.}
    \label{fig:F07Models}
\end{figure}
\begin{figure}
	\includegraphics[width=\columnwidth,angle=0]{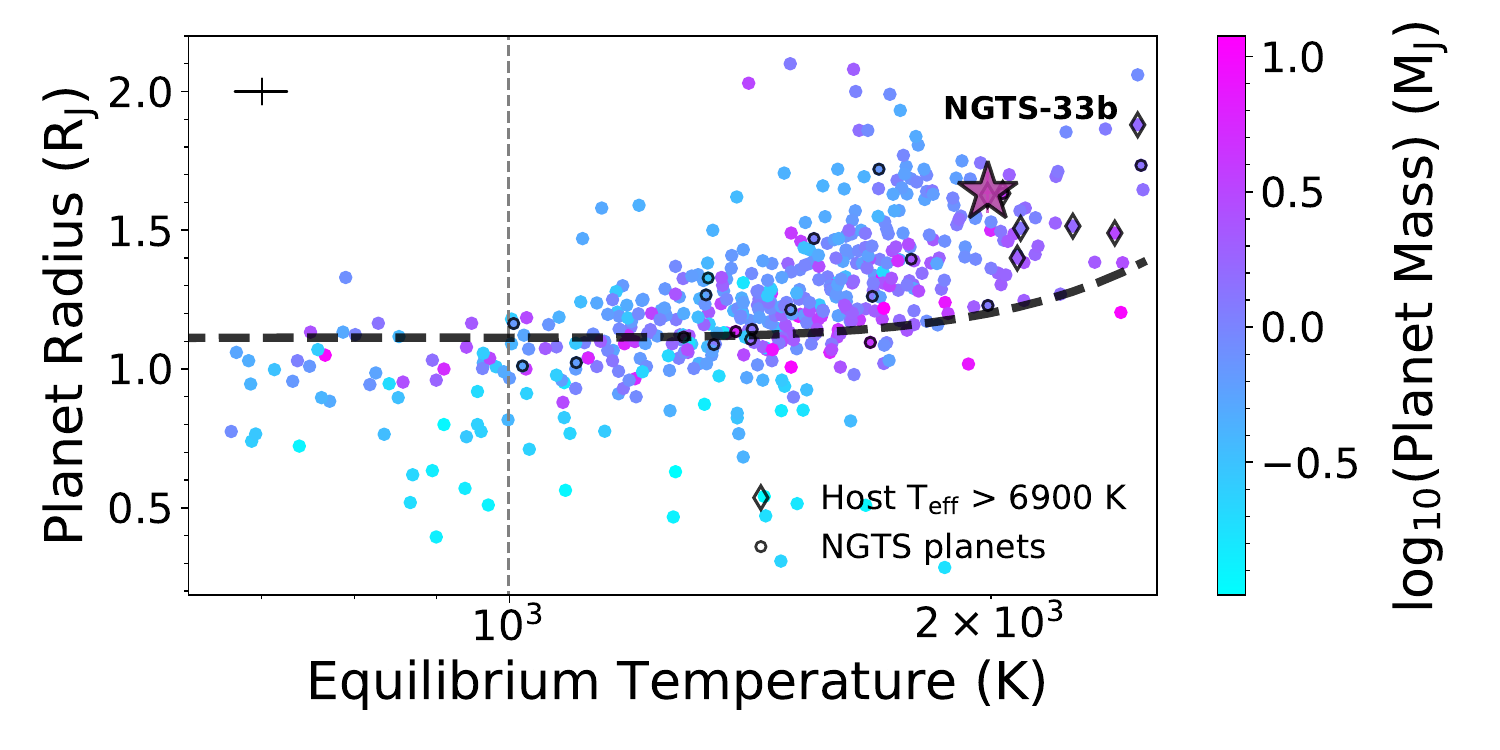}
	\includegraphics[width=\columnwidth,angle=0]{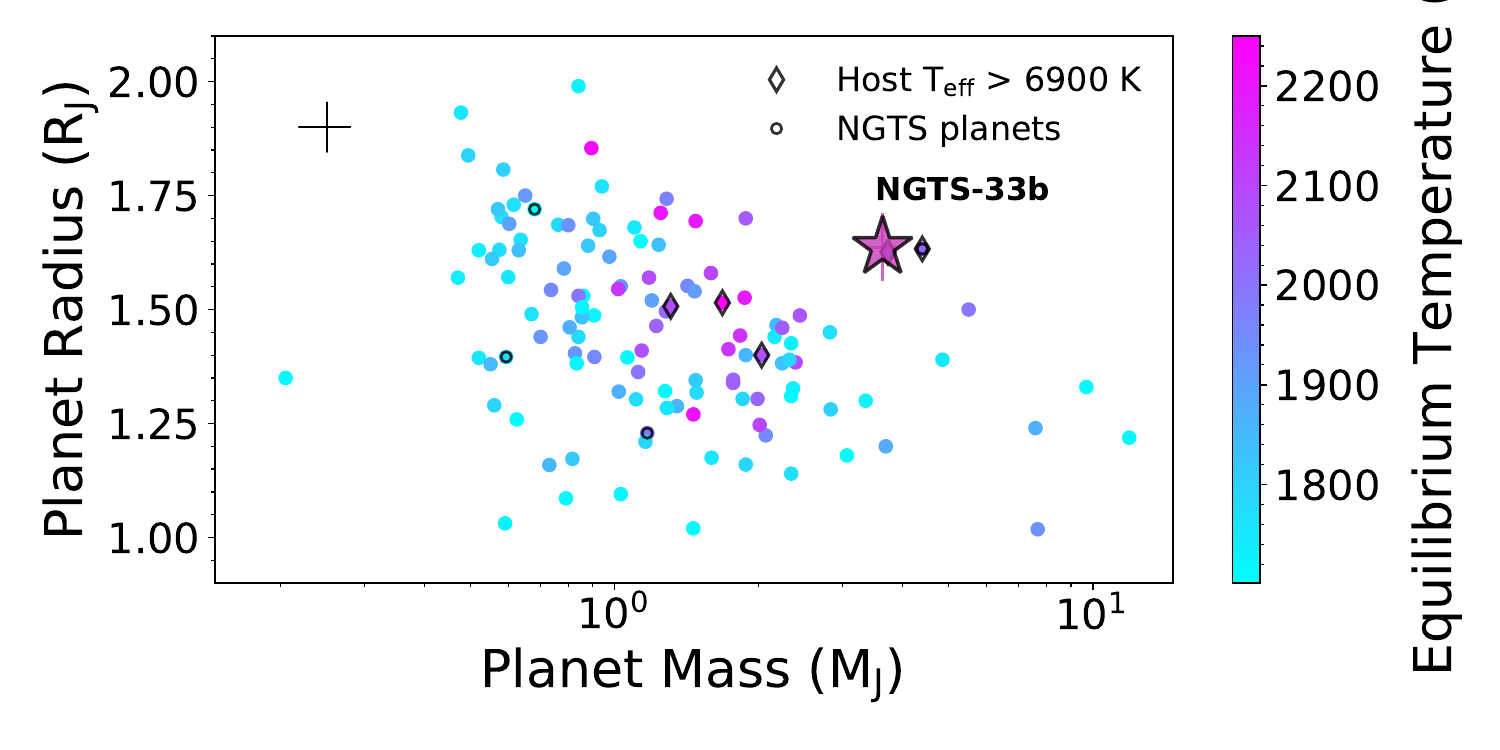}
    \caption{\textbf{Top}: Equilibrium temperature vs planet radius colour-coded by the logarithm of planet mass. Black dashed line represents an inflation-free model for a HJ evolved to 4.5 Gyr with a H/He composition adapted from \citet{thorngren2018bayesian}. \Nplanet\ is displayed by the red coloured star above the model near the image top right. Black cross at the top left corner represents the HJs parameters uncertainties standard deviation, and red open circles marks metal-poor K dwarf stars. \textbf{Bottom}: M$_{\rm p}$ vs R$_{\rm p}$ colour-coded by T$_{\rm eq}$ showing HJs from the top figure with T$_{\rm eq}$ between 1700-2300 K.}
    \label{fig:TeqVSRp}
\end{figure}
High incidence stellar flux into a planet's gaseous envelope seems to be the main driving mechanism responsible for the HJs radius anomaly. \citet{demory2011lack} estimates that an incident flux throughput of $\sim$ 2$\times10^5$ Wm${^{-2}}$ is required for planetary inflation to become observable. \citet{miller2011heavy} points out that the metallicity fraction (Z) also plays an important role, with higher Z fractions reducing the inflation efficiency. In addition, statistical analysis based on planetary thermal evolution models on a sample of 281 HJs \citep{thorngren2018bayesian} showed that the necessary conversion of incident flux into internal heating required to reproduce HJs observed radii, peak's at equilibrium temperature (T$_{\rm eq}) \sim$ 1500 K. Finally, \citet{hartman2016hat} shows that HJ radii grows as a function of main sequence star's fractional ages, i.e., as stars age on the main sequence, they brighten up, thus leading to higher planetary irradiation and hence higher T$_{\rm eq}$ of their orbiting planets. However, alternative scenarios exist to explain HJ radius anomalies, such as star-planet tidal interactions, which lead to internal heating of the planet, thus causing a radius inflation \citep[e.g., see][for a review]{fortney2021hot}.

We probed inflation levels from planetary structure models by \citet{fortney2007planetary} (hereafter F07). In Figure \ref{fig:F07Models} we show F07 models converted to the appropriate planet-star distance of $\sim 0.02$AU and evaluated at the model closest in age to \Nplanet, which shows that its measured radius is above the expected by $\sim 11\%$-$15\%$ assuming a core-free and 100 M$_{\oplus}$ core mass, respectively. Therefore, even at nearly $\sim50$ Myr, evolutionary models do not predict such radius, thus, despite its young age, stellar irradiation may be partially contributing to its large size. Had its age been $\sim$10 Myr, \Nplanet~radius would be nearly in agreement to F07 models, hence inflation-free, however, such young age is not corroborated by the Gyrochronology and cluster membership analysis (see $\S$ \ref{subsub:Agestimation}), hence it is unlikely that the planet is younger than $\sim10$ Myr given the uncertainties in the models and the systems' parameters. Moreover, the absence of infrared excess in the SED fitting (Fig. \ref{fig:sed}) provides further evidence on \Nstar~age lower limit as disks tend to disperse around 5-10 Myrs \citep{hillenbrand2005observational}. Therefore, we assume an age lower limit about 10 Myr, and conclude that \Nplanet~large radius is likely attributed to a combination of its very young nature and stellar irradiation, causing a radius inflation of $\sim 13\%$ on average.

In Figure \ref{fig:TeqVSRp} top panel we compared \Nplanet~to the THJ population in the R$_{\rm p}$ vs T$_{\rm eq}$ parameter space colour-coded by the logarithm of their masses. The planet's location, shown as a purple star symbol, suggests it is inflated compared to the HJ population as well as its counterparts with similar temperatures ($1700 \leq$T$_{\rm eq}$ $\geq~2300$ K).
However, we point out that the comparison above is partially affected by the HJ population broad ages. Fig. \ref{fig:TeqVSRp} lower panel highlights the THJs in the T$_{\rm eq}$ range in the R$_{\rm p}$ vs M$_{\rm Jup}$ phase space, where \Nplanet~ stands out as a massive and inflated planet. Such result places \Nplanet~as one of the first young and inflated HJ detected, with only 4 THJs younger than 100 Myr reported\footnote{https://exoplanet.eu/} thus far (e.g., K2-33 b, TOI-942 b, CoRoT-20 b, and WASP-33 Ab). Therefore, its discovery will add to the small but growing population of young planets, and subsequent follow-up will help place constraints on the formation and evolution of massive and hot systems.
\subsection{Is \Nplanet~suitable for atmospheric follow-up with JWST?}
\label{sub:TSM-ESM}
The discovery of transiting planets allowed the investigation of their atmospheric abundances, with short period HJs remaining the best targets for atmospheric follow-up. Their high equilibrium temperature (T${\rm eq} \geq 1000$ K) and lower densities ($\rho_{\rm p} < 0.3$ gcm$^{-3}$), induces scale heights such that the chemistry of inner layers are brought up, thus allowing the investigation of a variate of species by the transmission spectroscopy technique \citep{seager2000theoretical}. On the other hand, the planet's secondary eclipses allows the inference of day-side temperature as well as the albedo/reflectivity through emission spectroscopy. Such techniques have been vastly applied by ground- (e.g., ESPRESSO/VLT, HIRES/Keck) and space-based missions such as the Hubble Space Telescope, Spitzer, and recently, the JWST. 

\citet{stevenson2016transiting} selected 12 objects for JWST follow-up dubbed community targets (CT) based mostly on their properties, where optimal candidates are often quiet ($\log{R^{'}_{\rm HK}} \leq -4.8$), short period (P < 10 days), bright (J < 10.5 mag) and have orbital solutions and masses well-constrained. A high ecliptic latitude (b > 45$^{\circ}$) is also important as such targets being near or within the continuing viewing zone allow for multiple visits, consequently a higher SNR would be reached. Since \Nplanet~ matches several of these criteria, we computed its signal size per scale height defined in \citet{kempton2018framework} by assuming a cloud-free atmosphere and constant T$_{\rm eq}$, which is given by
\begin{equation}
\label{eq:signal-size}
    \Delta D = \frac{2k_{\rm B}T_{\rm eq}R_{\rm p}}{\mu gR^{2}_{\rm S}}    
\end{equation}
with k$_{\rm B}$, $\mu$, and g being the Boltzmann constant, mean molecular weight and the planet's gravity. From Eq. \ref{eq:signal-size} we estimated \Nplanet~$\Delta D$ of $\sim$ 53 ppm/H, which is comparable to the lower limit of the community target signal size range of 60-240 ppm/H, thus making \Nplanet~an interesting target for follow-up observations.

\begin{figure}
	\includegraphics[width=\columnwidth,angle=0]{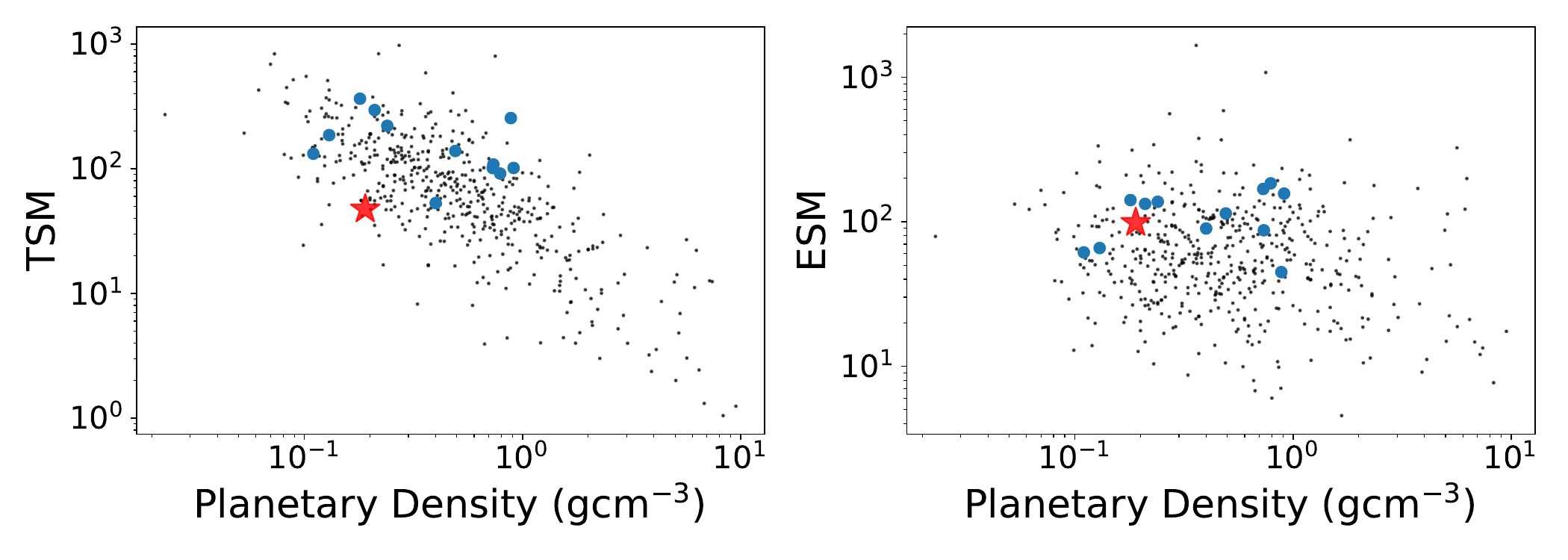}
        \includegraphics[width=\columnwidth,angle=0]{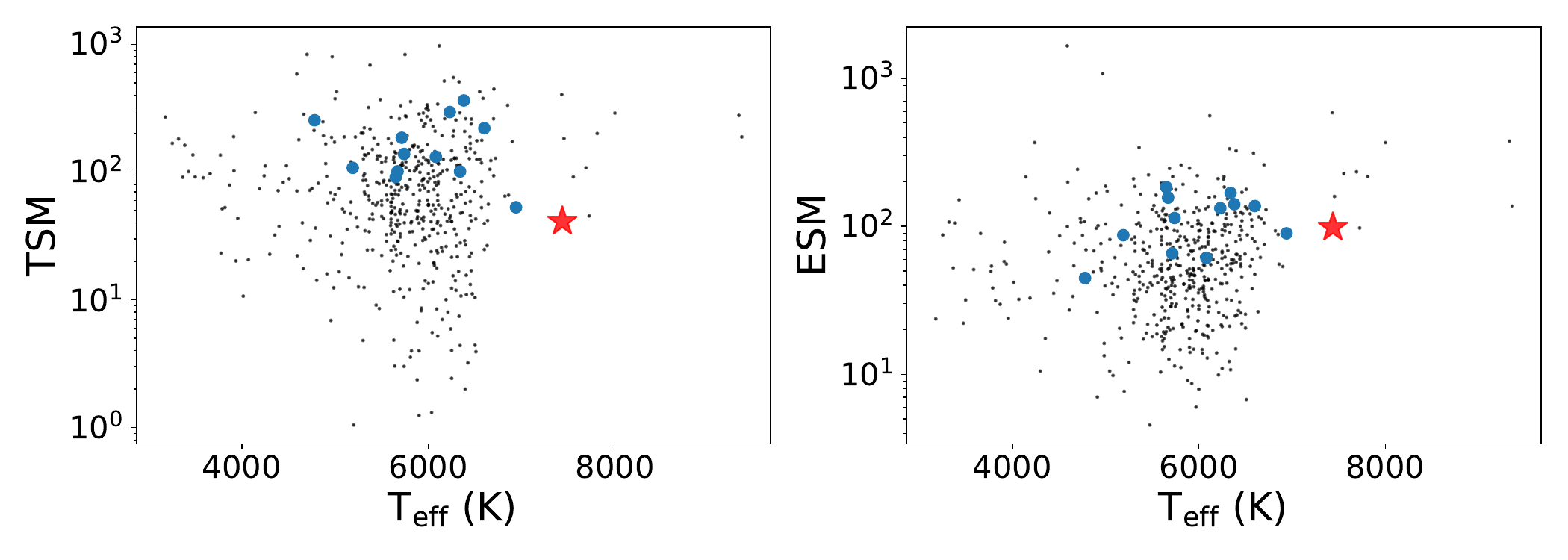}
    \caption{{\bf Top}: Transmission and emission spectroscopy metrics as a function of planetary density for the transiting HJ sample are shown in black circles, whereas JWST community targets by \citep{stevenson2016transiting} are marked by blue circles. \Nplanet~ is represented by the big red star. The TSM and ESM were computed homogeneously using the systems properties from the TEPCat catalogue.
    {\bf Bottom}: TSM and ESM as a function of host effective temperature. Colour schemes are the same as described in the top panel.}
    \label{fig:TSM-ESM}
\end{figure}

We have also computed the transmission and emission spectroscopy metrics TMS and EMS homogeneously from Eq. 1 and Eq. 4 by \citep{kempton2018framework}, shown in Fig. \ref{fig:TSM-ESM}. Although \Nplanet~seems less favorable for transmission spectroscopy as compared to the HJs and CT targets at similar densities (top left panel), its host sits in an under populated part of the TSM vs T$_{\rm eff}$ parameter space (bottom left panel), thus making the system an interesting candidate for atmosphere characterisation. In addition, the hosts with T$_{\rm eff} > 6900$ K, mostly left of \Nstar~are orbited by denser planets ($\rho_{\rm p} > 0.28$ gcm$^{-3}$), hence distinct atmospheric properties/abundances given their smaller scale heights. The ESM though indicates that \Nplanet~is an interesting target for JWST emission spectroscopy follow-up as its ESM is equivalent to CT and above most of HJ ESMs (top right panel). Similarly to the TSM, ESM vs T$_{\rm eff}$ shows that very few stars populate the hot edge (T$_{\rm eff} > 6900$ K) of the parameter space. Even though a handful of them ($\sim$ 4 out of 11) have been targeted for follow-up, the majority lacks atmospheric characterisation. Therefore, such part of the parameter space remains a golden region for an in-depth JWST atmospheric follow-up. 

\subsection{\Nplanet~obliquity follow-up}
\label{sec:RMfollowup}
When a transiting planet blocks part of the stellar disc, the spectral line centroids can be shifted, leading to an RV anomaly, well-known as the Rossiter-McLaughlin effect \citep[RM;
][]{triaud2009rossiter}. Such phenomenon allows us to measure the obliquity angle between the planet's orbital axis to that of the star. In fact, the technique was responsible to reveal that a handful of close-in massive giants have non-zero obliquities (e.g., KELT-17b, MASCARA-1b, TOI-1431b), with some even in retrograde orbits \citep[HAT-P-14b;][]{winn2011orbital}. In addition \citet{winn2011orbital} points that cool hosts tend to have planets with low obliquities, compared to hot hosts, hence indicating that even THJs may have undergone distinct migration channels, yet non-zero obliquities could have also been a result of primordial misalignment between the central star and its protoplanetary disc \citep{albrecht2022stellar}. Therefore, measuring the obliquity of \Nplanet~might provide valuable information on massive planetary systems evolution history.

The relatively large star-to-planet radius ratio and the host fast spinning rate indicates that \Nstar~is a good candidate for in-transit RV monitoring. The RM amplitude can be estimated by
\begin{equation}
A_{\rm RM} = \frac{2}{3}~\left(\frac{R_p}{R_{*}}\right)^{2}~v\sin{i_\ast}~\sqrt{1-b^{2}}    
\end{equation}
where b is the planet's impact parameter. An A$_{\rm RM} \sim$660.56 ms$^{-1}$ is expected for the planet, assuming $v\sin{i_\ast}$ $\sim$ 111.87 \kms from P$_{\rm rot}$ and R$_{\rm p}$. Such value is $\sim$ 1.7 higher than the planet induced RVs, which is explained by the relatively large coverage of the stellar disc by the planet, but mainly due to the star's extremely short rotation period of 0.67 days. In addition, the system's obliquity could possibly be measured directly through Doppler Doppler tomography \citep{cameron2010line, watanabe2022nodal}. Finally, \citet{gaudi2007prospects} showed that, for small planets of mass M on an edge-on circular orbit with period P, the ratio between the RM amplitudes and that of the planet induced velocity can be approximated by
\begin{equation}
\frac{A_{\rm RM}}{K} \sim0.3~(\frac{M}{M_{\rm Jup}})^{-1/3}~(\frac{P}{{\rm 3~days}})^{1/3}~(\frac{v\sin{i_\ast}}{5~{\rm km~s^{-1}}})    
\end{equation}
Using our derived values for M, P and $v\sin{i_\ast}$, we would expect an even larger ratio of about 4.3. Nonetheless, \Nplanet's RM effect is expected to be above the planet's induced velocity, thus it can easily be measured with high-resolution spectrographs (R $\geq$ 60,000).
\section{Conclusion}
\label{sec:concl}
We report the discovery of \Nplanet, a super-Jupiter with mass, radius, and bulk density of 3.63 $\pm$ 0.27 $\mjup$, 1.64 $\pm$ 0.07 $\rjup$ and 0.19 $\pm$ 0.03 \gccc, respectively. The planet orbits a massive A9V star every 2.83 days, whose mass, radius and effective temperature are of 1.60 $\pm$ 0.11 $\mstar$, 1.47 $\pm$ 0.06 \rstar, and 7437 $\pm$ 72 K, respectively. \Nstar~is one of the youngest hosts discovery thus far, having an age of 10-50 Myr, and the 5$^{th}$ hottest star hosting a THJ. In addition, membership analysis indicates that \Nstar~may be part of the Vela OB2 association. In fact, \Nstar~is not only young but also rare, laying amongst the 2$\%$ most massive THJ hosts, whereas the planet represents the $\sim$ 7$\%$ most massive THJs currently detected. Moreover, planetary structure models show that \Nplanet's radius is likely inflated by up to $\sim$ 15$\%$. Such is evidenced by its extremely low density compared to THJ of similar M and T$_{\rm eq}$. In fact, \Nplanet's bulk density is nearly 13$\%$ lower than that expected for its mass, thus likely a combination of its large radius due to youth and its irradiated atmosphere. In addition, we noticed what may be three populations of THJs distinguished by their T$_{\rm eq}$, with boundaries at T$_{\rm eq}$ $\sim$ 1400 K and $\sim$ 2300 K. Giants with T$_{\rm eq}$ < 1400 K were found to have larger $\rho_{\rm p}$ independent of mass, thus implying that their $\rho_{\rm p}$ are, to first order, a function of T$_{\rm eq}$ regardless of thermal evolution processes. Empirical linear models fitted to cooler (T$_{\rm eq}$<1400 K) and warmer ($1400 \leq$T$_{\rm eq}$ <2300 K) giants populations showed a difference in $\rho_{\rm p}$ of about 0.18, 0.23 and 0.62 \gccc for giants with masses of 0.6, 1 and 10 \mjup~planets, respectively.

The ESM computed for \Nplanet~puts it near the JWST community targets boundary, and one of the lowest in density amongst hot hosts, thus making it a suitable candidate for emission spectroscopy with JWST. Additionally, \Nstar's fast spin combined with the planet's large radius, gives it an expected RM amplitude of $\sim$ 660.56 \ms~, i.e. 1.7 times that of the RV induced variation generated by the planet, thus making it a good candidate for obliquity studies also.

Finally, with only 11 detections thus far, the number of massive Jupiters hosted by hot stars (T$_{\rm eff} \geq 6900$K) is strikingly scarce, thus the discovery of \Nplanet~will significantly add to the small but increasing population of massive THJs, which will help place further constraints on current formation and evolution models for such planetary systems.

\section*{Acknowledgements}
Based on data collected under the NGTS project at the ESO La Silla Paranal Observatory. The NGTS facility is operated by the consortium institutes with support from the UK Science and Technology Facilities Council (STFC) under projects ST/M001962/1, ST/S002642/1 and ST/W003163/1. 
This study is based on observations collected at the European Southern Observatory under ESO programme 105.20G9.
DRA acknowledges support of ANID-PFCHA/Doctorado Nacional-21200343, Chile, and thank the anonymous referee for the useful comments that improved the quality of this work.
JSJ greatfully acknowledges support by FONDECYT grant 1240738 and from the ANID BASAL projects ACE210002 and FB210003.
JIV acknowledges support of CONICYT-PFCHA/Doctorado Nacional-21191829.
Contributions at the University of Geneva by ML, FB and SU were carried out within the framework of the National Centre for Competence in Research "PlanetS" supported by the Swiss National Science Foundation (SNSF).
The contributions at the University of Warwick by PJW, SG, DB and RGW have been supported by STFC through consolidated grants ST/P000495/1 and ST/T000406/1.
The contributions at the University of Leicester by MGW and MRB have been supported by STFC through consolidated grant ST/N000757/1.

CAW acknowledges support from the STFC grant ST/P000312/1.
TL was also supported by STFC studentship 1226157.
MNG acknowledges support from the European Space Agency (ESA) as an ESA Research Fellow.
This project has received funding from the European Research Council (ERC) under the European Union's Horizon 2020 research and innovation programme (grant agreement No 681601).
The research leading to these results has received funding from the European Research Council under the European Union's Seventh Framework Programme (FP/2007-2013) / ERC Grant Agreement n. 320964 (WDTracer).
The contributions of ML and MB have been carried out within the framework of the NCCR PlanetS supported by the Swiss National Science Foundation under grants 51NF40\_182901 and 51NF40\_205606. ML also acknowledges support of the Swiss National Science Foundation under grant number PCEFP2\_194576.

\section*{DATA AVAILABILITY}

The data underlying this article are made available in its online supplementary material.




\bibliographystyle{mnras}
\bibliography{paper} 

\clearpage

\appendix
\section{Extra tables and figures}
\subsection{\texttt{ARIADNE} priors for the stellar characterisation}
\begin{table*}
	\centering
	\caption{\Nstar{} \texttt{ARIADNE} priors}
	\label{tab:priors-ariadne}
	\tabcolsep=0.11cm
	\begin{tabular}{ccc} 
Parameters		&	Prior & Hyperparameters	\\
\hline
\teff       &   Normal &  (7434,$100^2$)\\
\logg       &   Normal & (4.3, $0.4^2$)\\
\met        &   Normal & (-0.13, $0.23^2$)\\
Distance    &   Normal &  (438,$10^2$)\\
\rstar      &   Normal &  (1.48, $0.5^2$)\\
A$_{\rm V}$ &   Uniform & (0.00, 1.02)\\
\hline
	\end{tabular}
\end{table*}
%
\subsection{Global Modelling: Activity Signal}
\begin{table*}
	\centering
	\caption{Properties for the second Keplerian}
	\begin{tabular}{lc} 
	Property	&	Value \\
	\hline
    P (days) & 1.391558 (fixed) \\
    K (\ms) 	& 176$\pm$54        \\
    e 			& 0.0 (fixed)   \\
    $\omega~(\deg)$ &  90 (fixed) \\
    ${\rm M_o}~(\deg)$ & 64 (fixed) 	\\
	\hline
	\end{tabular}
    \label{tab:Kep2}
\end{table*}
\begin{figure*}
	\includegraphics[width=\columnwidth]{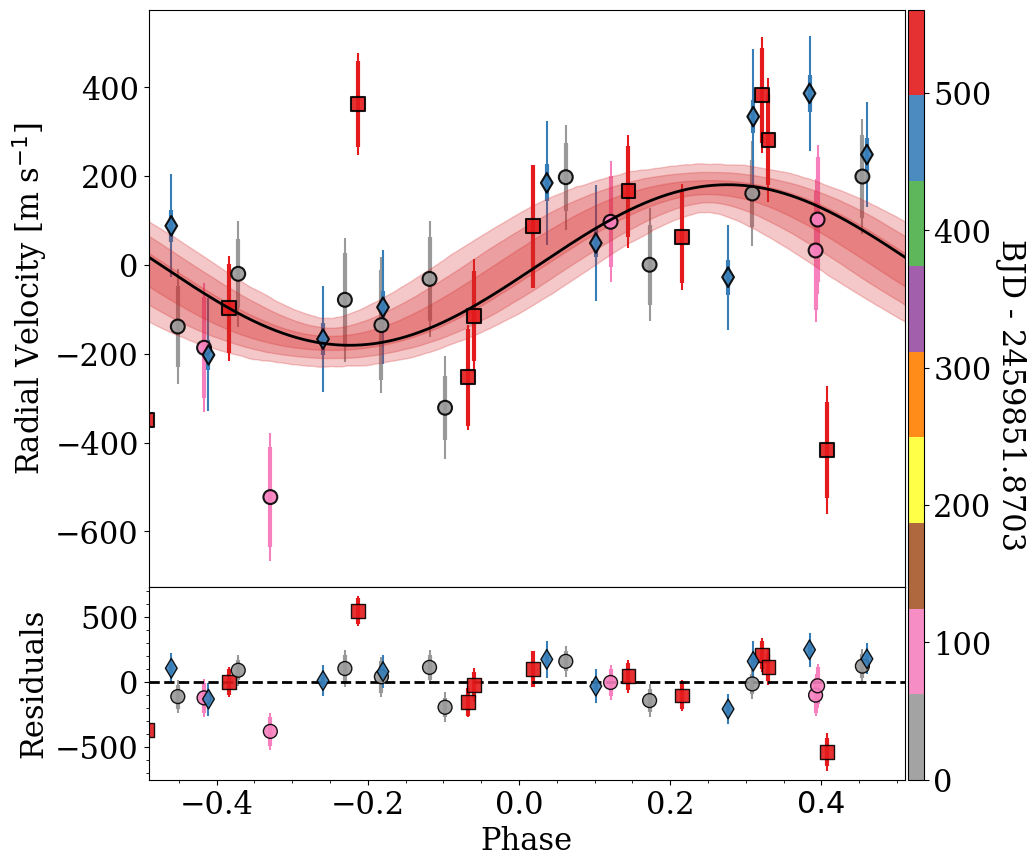}
    \caption{\textbf{Top}: RV phase-folded to the second Keplerian values associated to the stellar activity signal with fixed values from Table \ref{tab:Kep2}. \textbf{Bottom}: residuals to the best fit model.} 
    \label{fig:rvsKep2}
\end{figure*}
\subsection{Transit Timing Variation data}
\begin{table*}
	\centering
	\caption{\Nstar{} TTV data}
	\label{tab:ttvs}
	\tabcolsep=0.11cm
	\begin{tabular}{ccccc} 
Time (days; BJD$_{\rm TDB}$)	&	OC (minutes)       &{\rm low error} & {\rm upper error} & Pipeline\\
\hline
2458493.239828 & 1.15399 & 1.54222 & 1.45155 & TESS-SPOC \\
2458496.067799 & -1.01492 & 1.92430 & 1.91326 & TESS-SPOC \\
2458498.895770 & -1.82914 & 1.89144 & 2.20077 & TESS-SPOC \\
2458501.723741 & -0.34039 & 1.57699 & 1.59344 & TESS-SPOC \\
2458504.551712 & 2.08556 & 2.49663 & 2.06386 & TESS-SPOC \\
2458507.379683 & -1.10107 & 1.74482 & 2.02213 & TESS-SPOC \\
2458513.035625 & -0.60221 & 1.51767 & 1.54191 & TESS-SPOC \\
2458515.863596 & 1.61295 & 1.56667 & 1.47080 & TESS-SPOC \\
2459225.684317 & 0.40665 & 1.20163 & 1.24230 & TESS-SPOC \\
2459228.512288 & 0.68259 & 1.23309 & 1.13504 & TESS-SPOC \\
2459231.340259 & -1.48732 & 1.21970 & 1.25541 & TESS-SPOC \\
2459234.168230 & -0.43365 & 1.24079 & 1.25539 & TESS-SPOC \\
2459236.996201 & 0.57253 & 1.31606 & 1.35225 & TESS-SPOC \\
2459239.824172 & 0.69138 & 1.20454 & 1.23394 & TESS-SPOC \\
2459242.652143 & 0.18517 & 1.17847 & 1.20931 & TESS-SPOC \\
2459245.480114 & -0.40457 & 1.36060 & 1.30785 & TESS-SPOC \\
2459248.308085 & -1.17481 & 1.18139 & 1.15836 & TESS-SPOC \\
2459253.964027 & 0.78740 & 1.18390 & 1.18485 & TESS-SPOC \\
2459256.791998 & 0.26238 & 1.21833 & 1.27035 & TESS-SPOC \\
2459259.619969 & -0.97629 & 1.18017 & 1.19100 & TESS-SPOC \\
2459262.447940 & 0.12103 & 1.18787 & 1.19200 & TESS-SPOC \\
2459268.103882 & -0.57445 & 1.21199 & 1.24292 & TESS-SPOC \\
2459270.931853 & -0.51148 & 1.18199 & 1.20794 & TESS-SPOC \\
2459273.759824 & 0.11163 & 1.21530 & 1.17389 & TESS-SPOC \\
2459276.587795 & 0.78147 & 1.26831 & 1.24836 & TESS-SPOC \\
\hline
2459963.784748 & 0.52137 & 1.14355 & 1.13239 & QLP \\
2459966.612719 & -1.54366 & 1.24778 & 1.24212 & QLP \\
2459969.440690 & 0.10603 & 1.09343 & 1.04511 & QLP \\
2459972.268661 & 0.93827 & 1.32315 & 1.32652 & QLP \\
2459977.924603 & 1.11141 & 1.45005 & 1.39217 & QLP \\
2459980.752574 & -0.13346 & 1.04637 & 1.07595 & QLP \\
2459983.580545 & -0.80981 & 0.97173 & 0.97677 & QLP \\
2459986.408516 & 0.51174 & 1.00639 & 1.01724 & QLP \\
\hline
2458846.736203 & 0.843355 & 1.267368 & 1.241354& NGTS \\
2458880.671855 & -0.571765 & 1.229419 & 1.120578& NGTS \\
2458897.639681 & -0.649302 & 1.378999 & 1.412332& NGTS \\
2458914.607507 & 0.060029 & 1.175076 & 1.212128& NGTS \\
2459174.780839 & -0.262582 & 1.045301 & 1.142524& NGTS \\
2459208.716491 & 0.745968 & 1.103477 & 1.067075& NGTS \\
2459225.684317 & -0.402598 & 1.236785 & 1.264854& NGTS \\
	\end{tabular}
\end{table*}
\subsection{Spot-crossing events on \Nstar~and periodicity analysis on nearby object UCAC4 271-014742}
\begin{figure*}
	\includegraphics[width=\columnwidth]{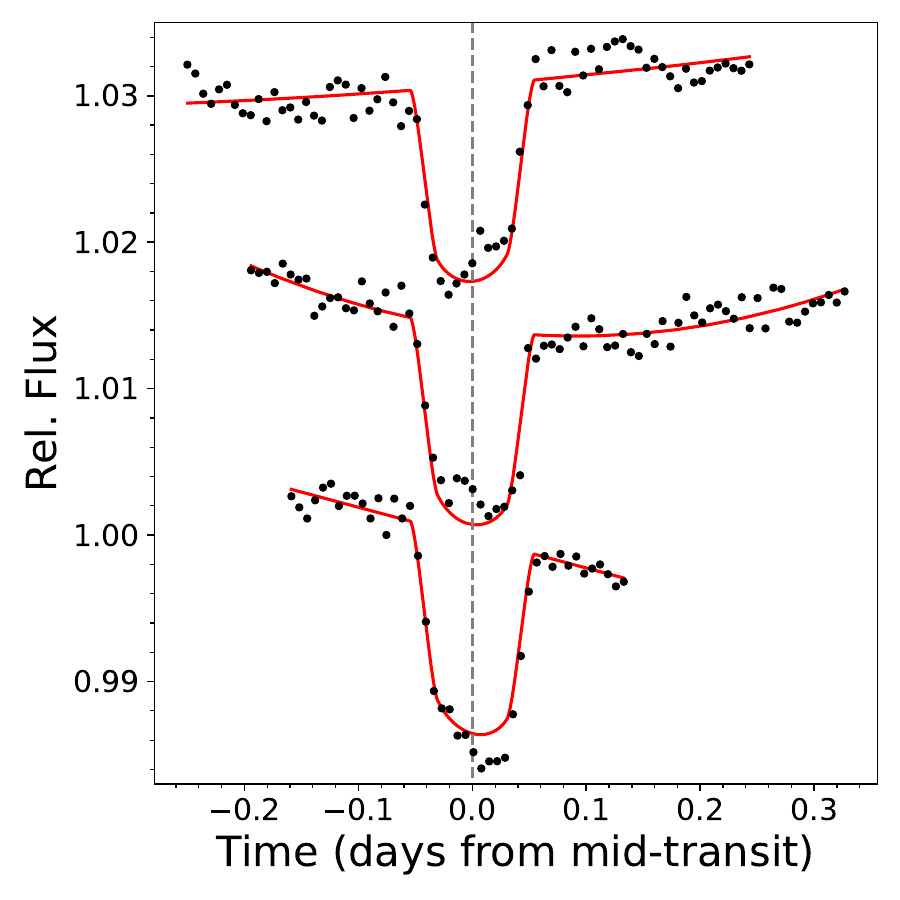}
    \caption{TESS-SPOC non-detrended time-series (in black) showing 3 potential spot-crossing events and stellar activity. A polynomial function was added to the transit model for visualisation purposes (in red). QLP and CDIPS lightcurves were visually check, thus roughly showing the same spot-crossing patterns. Time flows from top to bottom, with the first two events from sector 33, while the transit at the bottom being from sector 34.} 
    \label{fig:spot-crossing}
\end{figure*}
\begin{figure*}
\includegraphics[width=2\columnwidth,angle=0]{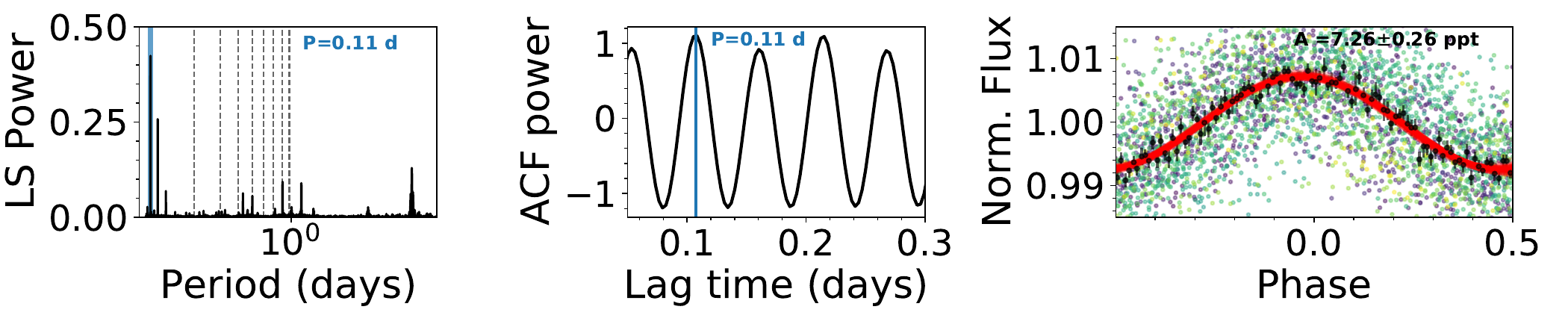}
    \caption{Rotation analysis on UCAC4 271-014742, a known pulsating variable star of V = 13.9 mag, period of 0.097174 day from GAIA, and 32" away from \Nstar. See Figure \ref{fig:rotation} for comparison and $\S$ \ref{sub:globalmodeling} and \ref{sub:rotation} for more information.} 
    \label{fig:UCAC4_contaminant}
\end{figure*}
%


\bsp	
\label{lastpage}
\end{document}